\documentclass[12pt]{article}
\pdfoutput=1
\usepackage{jheppub}
\usepackage{epsfig}
\usepackage{amsmath}
\usepackage{amssymb}
\usepackage{amsfonts}
\usepackage{amsxtra}
\usepackage{amsthm}
\usepackage{mathrsfs}
\usepackage{makeidx}
\usepackage{graphics}
\usepackage{dsfont}
\usepackage{mathtools}
\usepackage{graphicx}
\usepackage{subcaption}
\usepackage{placeins}
\usepackage{bm}
\usepackage[capitalise]{cleveref}
\usepackage{empheq}
\usepackage{colortbl}
\usepackage{xcolor}
\usepackage{enumerate}
\usepackage{titlesec}
\usepackage{longtable}
\usepackage{float}
\usepackage{color}
\usepackage{tikz}
\usepackage{xfrac}
\usepackage{footnote}
\usepackage{rotating}
\usepackage{lscape}
\usepackage{makecell}
\usepackage{environ}
\usepackage{tabularx}
\usepackage{subfiles}
\usepackage[export]{adjustbox}
\usepackage{ytableau}
\usepackage{tikz-3dplot}
\usepackage{slashed}
\usepackage{pifont}
\usepackage{multirow}
\usepackage{mdframed}
\usepackage{bbm}
\usepackage{ulem}

\usetikzlibrary{positioning,trees,decorations.pathmorphing,decorations.markings,decorations.pathreplacing,calc,shapes,patterns,arrows,chains,arrows.meta,fit,fadings,decorations.markings,graphs,graphs.standard,quotes}

\titleformat{\paragraph}
{\normalfont\normalsize\bfseries}{\theparagraph}{1em}{}
\titlespacing*{\paragraph}{0pt}{3.25ex plus 1ex minus .2ex}{1.5ex plus .2ex}

\def\CC{{\mathbb{C}}}

\theoremstyle{definition}

\catcode`\@=11   
\newdimen\@rotdimen
\newbox\@rotbox  

\def\@vspec#1{\special{ps:#1}}
\def\@rotstart#1{\@vspec{gsave currentpoint currentpoint translate
   #1 neg exch neg exch translate}}
\def\@rotfinish{\@vspec{currentpoint grestore moveto}}
\def\@rotr#1{\@rotdimen=\ht#1\advance\@rotdimen by\dp#1%
   \hbox to\@rotdimen{\hskip\ht#1\vbox to\wd#1{\@rotstart{90 rotate}%
   \box#1\vss}\hss}\@rotfinish}
\def\@rotl#1{\@rotdimen=\ht#1\advance\@rotdimen by\dp#1%
   \hbox to\@rotdimen{\vbox to\wd#1{\vskip\wd#1\@rotstart{270 rotate}%
   \box#1\vss}\hss}\@rotfinish}%
\def\@rotu#1{\@rotdimen=\ht#1\advance\@rotdimen by\dp#1%
   \hbox to\wd#1{\hskip\wd#1\vbox to\@rotdimen{\vskip\@rotdimen
   \@rotstart{-1 dup scale}\box#1\vss}\hss}\@rotfinish}%
\def\@rotf#1{\hbox to\wd#1{\hskip\wd#1\@rotstart{-1 1 scale}%
   \box#1\hss}\@rotfinish}%
\def\rotate{\@ifnextchar[{\@rotate}{\@rotate[l\right]}}
\def\@rotate[#1]#2{\setbox\@rotbox=\hbox{#2}\@nameuse{@rot#1}\@rotbox}

\catcode`\@=12

\usetikzlibrary{positioning}
\usetikzlibrary{chains}
\usetikzlibrary{arrows, arrows.meta ,fit,decorations.pathreplacing}
\tikzstyle{every picture}+=[remember picture]
\tikzstyle{na} = [baseline]

\usetikzlibrary{arrows, decorations.markings, calc, fadings, decorations.pathreplacing, patterns, decorations.pathmorphing, positioning}
\tikzset{>={Latex[width=1.5mm,length=1.5mm]}}
\usetikzlibrary{graphs,graphs.standard,quotes}

\def\node#1#2{\overset{#1}{\underset{#2}{{\color{gray} \bullet}}}}

\def\node#1#2{\overset{#1}{\underset{#2}{\circ}}}

\tikzstyle{every picture}+=[remember picture]
\tikzstyle{na} = [baseline=-.5ex]

\newcommand{\eg}{e.g.}

\newcommand{\ie}{i.e.}

\numberwithin{equation}{section}
\newcommand{\bes}[1]{\begin{equation} \begin{split} #1\end{split} \end{equation}}

\newcommand{\nn}{\nonumber}

\newcommand{\be}{\begin{equation}} \newcommand{\ee}{\end{equation}}
\newcommand{\bea}{\begin{equation} \begin{aligned}} \newcommand{\eea}{\end{aligned} \end{equation}}

\def\tilde{\widetilde}

\def\hat{\widehat}

\def\bar{\overline}

\def\rt2{\sqrt{2}}

\def\mod{{\rm mod}}

\def\Tr{\mathop{\rm Tr}}

\def\CB{{\cal B}}
\def\CC{{\cal C}}

\def\CG{{\cal G}}

\def\CI{{\cal I}}

\def\CM{{\cal M}}
\def\CN{{\cal N}}

\def\CQ{{\cal Q}}

\def\CS{{\cal S}}

\def\CZ{{\cal Z}}

\def\1{{\ds 1}}

\newcommand{\fg}{\mathfrak{g}}
\newcommand{\fm}{\mathfrak{m}}

\def\repa{\raise4pt\hbox{$\square$}\mkern-14mu\raise-4pt\hbox{$\square$}}
\def\repab{\overline{\raise4pt\hbox{$\square$}\mkern-14mu\raise-4pt\hbox{$\square$}\mkern-1mu}}

\def\smileface{\ensuremath{\hbox{\large$\bigcirc$}\mkern-15mu\raise-1pt\hbox{\scriptsize$\smallsmile$}%
\mkern-10mu\raise4pt\hbox{..}\mkern4mu}}
\def\frownface{\ensuremath{\hbox{\large$\bigcirc$}\mkern-15mu\raise-1pt\hbox{\scriptsize$\smallfrown$}%
\mkern-10mu\raise4pt\hbox{..}\mkern4mu}}

\newcommand{\ba}{\begin{array}}
\newcommand{\ea}{\end{array}}
\newcommand{\bi}{\begin{itemize}}
\newcommand{\ei}{\end{itemize}}
\def\vec#1{\bm{#1}}
\def\bea#1\eea{\allowdisplaybreaks \begin{align}#1\end{align}}
 \newcommand{\ben}{\begin{enumerate}}
\newcommand{\een}{\end{enumerate}}
\newcommand{\bean}{\begin{eqnarray*}}
\newcommand{\eean}{\end{eqnarray*}}
\newcommand{\eref}[1]{(\ref{#1})}

\newcommand{\tQ}{\widetilde{Q}}

\newcommand{\BC}{\mathbb{C}}

\newcommand{\BZ}{\mathbb{Z}}

\newcommand{\Sym}{\mathrm{Sym}}

\definecolor{light-gray}{gray}{0.5}

\newcommand{\blue}{\color{blue}}

\newcommand{\red}{\color{red}}

\newcommand{\udl}[1]{\mathrm{d} #1 \,}

\def\ga{\alpha}

\def\Gp{\Phi}

\def\aup#1 {\overset{#1}{\uparrow} \, \overset{\tilde{#1}}{\downarrow}}

\tikzset{snake it/.style={decorate, decoration={snake, amplitude=.4mm, segment length=2mm,
                       post length=0mm,pre length=0mm}}}
                       
 \newcommand{\GCD}{\mathrm{GCD}}

\hypersetup{
	pdftitle={Zero-form and one-form symmetries of the ABJ and related theories},    
	pdfauthor={\textcopyright\ Emanuele Beratto, Noppadol Mekareeya, Matteo Sacchi},     
	pdfsubject={hep-th},   
	pdfcreator={pdfLaTex},   
	pdfproducer={LaTex}, 
	pdfkeywords={},
	colorlinks=true,
}

\makeatletter
\newsavebox{\measure@tikzpicture}
\NewEnviron{scaletikzpicturetowidth}[1]{%
  \def\tikz@width{#1}%
  \begin{lrbox}{\measure@tikzpicture}%
  \BODY
  \end{lrbox}%
  \pgfmathparse{#1/\wd\measure@tikzpicture}%
  \BODY
}
\makeatother

\makeatletter
\def\squarecorner#1{
    \pgf@x=\the\wd\pgfnodeparttextbox%
    \pgfmathsetlength\pgf@xc{\pgfkeysvalueof{/pgf/inner xsep}}%
    \advance\pgf@x by 2\pgf@xc%
    \pgfmathsetlength\pgf@xb{\pgfkeysvalueof{/pgf/minimum width}}%
    \ifdim\pgf@x<\pgf@xb%
        \pgf@x=\pgf@xb%
    \fi%
    \pgf@y=\ht\pgfnodeparttextbox%
    \advance\pgf@y by\dp\pgfnodeparttextbox%
    \pgfmathsetlength\pgf@yc{\pgfkeysvalueof{/pgf/inner ysep}}%
    \advance\pgf@y by 2\pgf@yc%
    \pgfmathsetlength\pgf@yb{\pgfkeysvalueof{/pgf/minimum height}}%
    \ifdim\pgf@y<\pgf@yb%
        \pgf@y=\pgf@yb%
    \fi%
    \ifdim\pgf@x<\pgf@y%
        \pgf@x=\pgf@y%
    \else
        \pgf@y=\pgf@x%
    \fi
    \pgf@x=#1.5\pgf@x%
    \advance\pgf@x by.5\wd\pgfnodeparttextbox%
    \pgfmathsetlength\pgf@xa{\pgfkeysvalueof{/pgf/outer xsep}}%
    \advance\pgf@x by#1\pgf@xa%
    \pgf@y=#1.5\pgf@y%
    \advance\pgf@y by-.5\dp\pgfnodeparttextbox%
    \advance\pgf@y by.5\ht\pgfnodeparttextbox%
    \pgfmathsetlength\pgf@ya{\pgfkeysvalueof{/pgf/outer ysep}}%
    \advance\pgf@y by#1\pgf@ya%
}
\makeatother

\pgfdeclareshape{square}{
    \savedanchor\northeast{\squarecorner{}}
    \savedanchor\southwest{\squarecorner{-}}

    \foreach \x in {east,west} \foreach \y in {north,mid,base,south} {
        \inheritanchor[from=rectangle]{\y\space\x}
    }
    \foreach \x in {east,west,north,mid,base,south,center,text} {
        \inheritanchor[from=rectangle]{\x}
    }
    \inheritanchorborder[from=rectangle]
    \inheritbackgroundpath[from=rectangle]
}

\tikzset{stretch/.initial=1}
\newcommand\drawloop[4][]%
   {\draw[shorten <=0pt, shorten >=0pt,#1]
      ($(#2)!\pgfkeysvalueof{/tikz/stretch}!(#2.#3)$)
      let \p1=($(#2.center)!\pgfkeysvalueof{/tikz/stretch}!(#2.north)-(#2)$),
          \n1= {veclen(\x1,\y1)*sin(0.5*(#4-#3))/sin(0.5*(180-#4+#3))}
      in arc [start angle={#3-90}, end angle={#4+90}, radius=\n1]%
   }

\title{Zero-form and one-form symmetries of the ABJ and related theories}
\author[a,b]{Emanuele Beratto,} 
\author[b,c]{Noppadol Mekareeya,}
\author[a,b,d]{Matteo Sacchi}
\affiliation[a]{Dipartimento di Fisica, Universit\`a di Milano-Bicocca, \\ Piazza della Scienza 3, I-20126 Milano, Italy}
\affiliation[b]{INFN, sezione di Milano-Bicocca, \\Piazza della Scienza 3, I-20126 Milano, Italy}
\affiliation[c]{Department of Physics, Faculty of Science, \\
Chulalongkorn University, Phayathai Road, \\
Pathumwan, Bangkok 10330, Thailand}
\affiliation[d]{Mathematical Institute, University of Oxford, \\ Andrew-Wiles Building, Woodstock Road, \\
Oxford, OX2 6GG, United Kingdom}
\emailAdd{emanuele.beratto@gmail.com}
\emailAdd{n.mekareeya@gmail.com}
\emailAdd{matteo.sacchi@maths.ox.ac.uk}
\abstract{The zero-form and one-form global symmetries of the Aharony-Bergman-Jafferis (ABJ) and related theories, with at least $\CN=6$ supersymmetry in three dimensions, are examined in detail. Starting from well-known dualities between theories with orthogonal and symplectic gauge groups and those with unitary gauge groups, we gauge their one-form symmetries or their subgroups and obtain new dualities. One side of the latter involves theories with special orthogonal and symplectic gauge groups, and the other side involves theories with unitary gauge groups; there is a discrete quotient on one or both sides of the duality. We study the refined superconformal indices of such theories and map the symmetries across the dualities, with particular attention to their discrete part. As a generalisation, we also find a new duality between a circular quiver with a discrete quotient of alternating special orthogonal and symplectic gauge groups and a three-dimensional $\CN=4$ circular (Kronheimer-Nakajima) quiver with unitary gauge groups, whose Higgs or Coulomb branch describes an instanton on a singular orbifold.}
\begin{document}
\maketitle

\section{Introduction}
The Aharony-Bergman-Jafferis-Maldacena (ABJM) $U(N)_k \times U(N)_{-k}$ theories\footnote{In this paper, $G_k$ denotes gauge group $G$ with Chern-Simons level $k$.} \cite{Aharony:2008ug} and the Aharony-Bergman-Jafferis (ABJ) $U(N+x)_k \times U(N)_{-k}$ theories \cite{Aharony:2008gk} constitute a large class of three-dimensional superconformal field theories (SCFTs) with $\CN=6$ and in some special cases $\mathcal{N}=8$ supersymmetry.  As observed in \cite{Aharony:2008gk} and further studied in \cite{Cheon:2012be}, some of these theories are dual to the ABJ theories with orthogonal and symplectic gauge groups:
\bes{ \label{dualityknown}
O(2N)_2 \times USp(2N)_{-1} \,\, &\longleftrightarrow \,\, U(N)_4 \times U(N)_{-4} \\
O(2N+2)_2 \times USp(2N)_{-1} \,\, &\longleftrightarrow \,\, U(N+2)_4 \times U(N)_{-4} \\
O(2N+1)_2 \times USp(2N)_{-1} \,\, &\longleftrightarrow \,\, U(N+1)_4 \times U(N)_{-4}
}

In this paper, we revisit the theories and dualities in \eref{dualityknown} in the context of generalised global symmetries. In particular, we investigate one-form symmetries in such theories and gaugings thereof.  As pointed out in \cite{Aharony:2013hda, Gaiotto:2014kfa}, the study of higher-form symmetries and extended operators leads to new insight on several structures of the theory, especially distinctions between theories with the same gauge algebra but with different global structures of the gauge group. This turns out to be the case here. In particular, by gauging the one-form symmetries (or their subgroups) of theories in \eref{dualityknown}, we find the following {\bf new} dualities:
\bes{ \label{newSOeven}
SO(2N)_2 \times USp(2N)_{-1} \,\, &\longleftrightarrow \,\, [U(N)_4 \times U(N)_{-4}]/\BZ_2 \\
[SO(2N)_2 \times USp(2N)_{-1}]/\BZ_2  \,\, &\longleftrightarrow \,\, [U(N)_4 \times U(N)_{-4}]/\BZ_4 \\
SO(2N+2)_2 \times USp(2N)_{-1} \,\, &\longleftrightarrow \,\, [U(N+2)_4 \times U(N)_{-4}]/\BZ_2
}
We point out that the left hand side of each duality involves a special orthogonal gauge group.  The discrete zero-form and one-form global symmetries of each theory are studied in-depth in this paper.  We find that the matching of global symmetries across duality is rather intricate, \eg~ for the theories in the first line, the $\BZ_2$ zero-form charge conjugation symmetry of the left theory becomes a subgroup of the $U(1) \times \BZ_{\GCD(N,4)}$ zero-form symmetry of the right theory, where the latter was pointed out in \cite{Bergman:2020ifi}. However, not all global symmetries of the unitary theory are manifest in the (special) orthogonal symplectic theory.  We discuss this point, as well as the map of the symmetries across the duality, extensively in the main text of the paper.

Note that, in three spacetime dimensions, gauging of a one-form symmetry yields a zero-form symmetry, and vice-versa \cite{Gaiotto:2014kfa}.  In many cases, this observation allows us to indirectly study the one-form symmetry in the original theory via the zero-form symmetry in another theory where such a one-form symmetry is gauged.  The (refined) superconformal index \cite{Bhattacharya:2008zy,Bhattacharya:2008bja, Kim:2009wb,Imamura:2011su, Kapustin:2011jm, Dimofte:2011py} serves as a particularly useful tool for studying the zero-form symmetry of the latter and thus contains information about the one-form symmetry of the original theory. The superconformal index in 3d is also sensitive to the global structure of the gauge groups and this makes it powerful to determine the correct ones in order for dualities, such as those in \eqref{dualityknown} and \eqref{newSOeven}, to hold. Moreover, as we shall comment again momentarily, it can also be used to detect topological sectors, which usually bring along one-form symmetries, by turning on background magnetic fluxes for zero-form global symmetries. It was already observed in \cite{Eckhard:2019jgg} that the Witten index is an observable that in 3d is sensitive to the higher form symmetries of the theory. The superconformal index that we will use extensively in our analysis is a generalisation of the Witten index refined with fugacities for the $R$-symmetry and for the zero-form global symmetries.

We also study theories with odd (special) orthogonal and symplectic gauge groups.  We find that the following four theories have the {\it same} refined superconformal indices:
\bes{
O(2N+1)_{2} \times USp(2N)_{-1} &\qquad  U(N+1)_4\times U(N)_{-4} \\
SO(2N+1)_{2} \times USp(2N)_{-1} &\qquad  [U(N+1)_4\times U(N)_{-4}]/\BZ_2
}
where the duality of the theories in the first line is known (see the last line of \eref{dualityknown}) and the duality in the second line follows from the first line by gauging the one-form symmetry of the latter. The equality of these indices leads us to conclude that the one-form symmetry of the theories in the first line acts trivially on the spectrum of line operators, and the zero-form charge conjugation symmetry acts trivially on the theories in the second line.  The former conclusion leads to the proposal that a certain set of line operators is absent from the theories in the first line.

We generalise the above result by considering a circular quiver with alternating special orthogonal and symplectic gauge groups.  Specifically, we find the duality between the following two theories: (1) the $\BZ_2$ discrete quotient of the circular quiver with alternating $SO(2)_2$ and $USp(2)_{-1}$ gauge groups and bifundamental half-hypermultiplets between each pair of such groups, and (2) a 3d $\CN=4$ circular quiver with a collection of $U(1)$ gauge groups, bifundamental hypermultiplets between each pair of such gauge groups, and a hypermultiplet with charge 1 under each gauge group. The latter is also known as a Kronheimer--Nakajima quiver \cite{kronheimer1990yang} and its Higgs or Coulomb branch describes an instanton on a singular orbifold.

The paper is organised as follows.  In Section \ref{sec:gendiss}, we gauge one-form symmetries (or their subgroups) of the theories in the first two lines of \eref{dualityknown} in order to obtain the theories in \eref{newSOeven}.  Mixed anomalies and non-trivial extensions of zero-form and one-form symmetries of such theories are discussed in detail.  We also present some interesting special cases that relate these theories to other known theories, such as $\CN=8$ super-Yang-Mills. In Section \ref{sec:indices}, we study the superconformal indices of all of the aforementioned theories and match them across the dualities. We point out the symmetries that are not manifest, especially in the (special) orthogonal symplectic theories, and cannot be refined in the index.  For the unitary theories, we study the relevant and marginal operators in detail, as well as the extra supersymmetry currents leading to $\mathcal{N}=6$ or $\mathcal{N}=8$.  We conclude the paper and propose some directions for further study in Section \ref{sec:conclusions}.  The conventions for the superconformal index adopted in this paper are summarised in Appendix \ref{app:index}. In Appendix \ref{app:appetiser} we demonstrate, via an example of the duality appetiser \cite{Jafferis:2011ns}, that the superconformal index can be sensitive to the topological sector when the background magnetic flux for a zero-form global symmetry is turned on.

\subsection*{Notation and convention}
\bi
\item We denote the ABJM theory \cite{Aharony:2008ug} by $U(N)_k \times U(N)_{-k}$, where the four chiral multiplets in the bifundamental representation and the superpotential are implicit.  Similarly, the ABJ theories \cite{Aharony:2008gk} are denoted by $U(N+x)_k \times U(N)_k$ and $O(2N+x)_{2k} \times USp(2N)_{-k}$.  They can be subject to a discrete quotient and the orthogonal group can be replaced by a special orthogonal group, depending on the theory we are considering.   

We use a similar notation for a circular quiver, \eg~ $SO(2)_2 \times USp(2)_{-1} \times SO(2)_2 \times USp(2)_{-1}$ denotes a quiver gauge theory with four gauge groups and the chiral multiplets in the bifundamental representation of each $SO(2) \times USp(2)$ pair.
\item Whenever we would like to emphasise the type of symmetries, the (ordinary) zero-form global symmetry is denoted by a superscript $[0]$ and the one-form global symmetry is denoted by a superscript $[1]$. 
\item We denote by a subscript the type of global symmetries, for example, the zero-form charge conjugation symmetry is denoted by $(\BZ^{[0]}_2)_\CC$.  Unless stated otherwise, we label the symmetry according to the notation of the fugacity in the superconformal index, \eg~ the fugacity for the zero-form magnetic symmetry of an orthogonal group is $\zeta$, and so we denote such a symmetry by $(\BZ^{[0]}_2)_\zeta$.
\ei

\section{General discussion} \label{sec:gendiss}
One of the main objectives of this paper is to study the interrelation between the following dualities:
\begin{equation}
\begin{tikzpicture}[baseline]

\node (X) at (6,0) {$\longleftrightarrow$};
\node (Y) at (6,-1.3) {$\longleftrightarrow$};
\node (Z) at (6,-2.6) {$\longleftrightarrow$};

\node[anchor=west] (A) at (-1,0) {I: \quad $O(2N)_2 \times USp(2N)_{-1}$};
\node[anchor=west] (B) at (-1,-1.3) {II: \quad $SO(2N)_2 \times USp(2N)_{-1}$};
\node[anchor=west] (C) at (-1,-2.6) {III: $[SO(2N)_2 \times USp(2N)_{-1}]/\BZ_2$};
\draw[->] (5.5,-0.2) -- node [left] {\footnotesize $\mathbb{Z}_{2}^{[1]}$}  (5.5,-1.1);
\draw[->] (5.5,-1.5) --  node [left] {\footnotesize $\mathbb{Z}_{2}^{[1]}$} (5.5,-2.4);

\node[anchor=west] (P) at (7,0) {$U(N)_4 \times U(N)_{-4}$};
\node[anchor=west] (Q) at (7,-1.3) {$[U(N)_4 \times U(N)_{-4}]/\BZ_2$};
\node[anchor=west] (R) at (7,-2.6) {$[U(N)_4 \times U(N)_{-4}]/\BZ_4$};
\draw[<-] (6.5,-0.2) --node [right] {\footnotesize $\mathbb{Z}_{2}^{[0]}$} (6.5,-1.1);
\draw[<-] (6.5,-1.5) -- node [right] {\footnotesize $\mathbb{Z}_{2}^{[0]}$}(6.5,-2.4);
\end{tikzpicture}
\end{equation}
where a downwards arrow with the label $\BZ^{[1]}_2$ denotes the gauging of the $\BZ^{[1]}_2$ one-form symmetry and an upwards arrow with the label $\BZ^{[0]}_2$ denotes the gauging of the $\BZ^{[0]}_2$ zero-form symmetry.  For convenience, we respectively use (L) and (R) to denote the left and right descriptions of each duality I, II or III; for example, II(L) denotes the left description of the duality II.  To the best of our knowledge, dualities II and III are {\bf new}, whereas duality I was conjectured in \cite{Aharony:2008gk} and studied in more detail in \cite{Cheon:2012be}.

Let us discuss each theory in detail.  As pointed out in \cite[Section 2.1]{Bergman:2020ifi}, the theory I(R) has a $U(1)^{[0]}_{\text{top}}$ topological symmetry and a $\BZ_4^{[1]}$ one-form global symmetry, with a mixed anomaly characterised by the following short exact sequence:
\bes{
\text{I(R)}: \quad 0~\rightarrow~\BZ_4^{[1]}~\rightarrow~ \BZ_4^{[1]} \times U(1)^{[0]}_{\text{top}}~\rightarrow~U(1)^{[0]}_{\text{top}}~\rightarrow~  0
}
Note that there is no non-trivial extension between these two symmetries. On the other hand, the $U(1)^{[0]}_{\text{top}}$ is not manifest in the description I(L).  As explained in \cite[Sections 2.3 and 2.4]{Cordova:2017vab} and \cite[Section 6.2]{Hsin:2020nts}, the $\BZ_4^{[1]}$ one-form symmetry of the theory I(L) arises from a non-trivial extension between the centre $(\BZ^{[1]}_{2})_{\text{centre}}$ one-form symmetry\footnote{Both $O(2N)$ and $USp(2N)$ gauge groups have a $\BZ_2$ centre (see \eg~\cite[Table 3]{Cordova:2017vab})
. However, the half-hypermultiplets in the bifundamental representation screen the diagonal combination of $\BZ_2 \times \BZ_2$, and so we are left with one $\BZ_2$ centre symmetry, which we denoted by $(\BZ^{[1]}_{2})_{\text{centre}}$.} and the $(\BZ^{[1]}_{2})_{\hat{\CC}}$ one-form symmetry controlled by the dynamical gauge fields for the zero-form charge conjugation symmetry.  We denote the latter by $(\BZ^{[1]}_{2})_{\hat{\CC}}$.  This is characterised by the short exact sequence:
\bes{ \label{SESIL}
\text{I(L)}: \quad 0~\rightarrow~(\BZ^{[1]}_{2})_{\hat{\CC}}~\rightarrow~ \BZ_4^{[1]}~\rightarrow~(\BZ^{[1]}_{2})_{\text{centre}}~\rightarrow~  0
}

Let us now gauge the one-form symmetry in each theory in the duality I.  In the theory I(L), this is identified as gauging $(\BZ^{[1]}_{2})_{\hat{\CC}}$, whereas in the theory I(R), this corresponds to gauging a $\BZ_2^{[1]}$ subgroup of the $\BZ_4^{[1]}$ one-form symmetry.  As a result, we obtain a $\BZ_2^{[0]}$ zero-form symmetry in each description of the duality II.  In the description II(L) this is identified as the zero-form charge conjugation symmetry $(\BZ^{[0]}_{2})_\CC$, and in the description II(R) this is identified with the zero-form discrete topological symmetry, denoted by $(\BZ^{[0]}_2)_{g''}$,\footnote{This notation is chosen to be consistent with \eref{defwpgpp}.} arising from the $\BZ_2$ discrete gauging.  We thus have the correspondence
\bes{
(\BZ^{[0]}_{2})_\CC \quad \longleftrightarrow \quad (\BZ^{[0]}_{2})_{g''}~.
}
After gauging, each description of the duality II also has a $\BZ_4^{[1]}/(\BZ^{[1]}_{2})_{\hat \CC} = \BZ_{2}^{[1]}$ one-form symmetry.  As pointed out in \cite[Section 2.5]{Bergman:2020ifi}, the theory II(R) has a $U(1)^{[0]} \times \BZ^{[0]}_{\GCD(N,2)}$ zero-form symmetry.  Similarly to \cite[(2.18)]{Bergman:2020ifi}, the latter can be viewed as a non-trivial extension of the $U(1)^{[0]}_{\text{top}}$ topological symmetry and the zero-form charge conjugation symmetry $(\BZ^{[0]}_{2})_{g''}$. This is characterised by the short exact sequence:
\bes{ \label{SESIIR}
\text{II(R)}: \quad 0~\rightarrow~(\BZ^{[0]}_{2})_{g''}~\rightarrow~ U(1)^{[0]} \times \BZ^{[0]}_{\GCD(N,2)}~\rightarrow~U(1)^{[0]}_{\text{top}}~\rightarrow~  0~.
}
Following the discussion in \cite{Tachikawa:2017gyf}, if we start from \eref{SESIL} and gauge the $(\BZ^{[1]}_{2})_{\hat{\CC}}$ one-form symmetry, we obtain
\bes{ \label{SESIIL}
\text{II(L)}: \quad 0~\rightarrow~(\BZ^{[0]}_{2})_{\CC} ~\rightarrow~  (\BZ^{[0]}_{2})_{\CC} \times (\BZ^{[1]}_{2})_{\text{centre}} ~\rightarrow~(\BZ^{[1]}_{2})_{\text{centre}}~\rightarrow~  0
}
with a mixed anomaly between the zero-form charge conjugation symmetry $(\BZ^{[0]}_{2})_{\CC}$ and the centre $\BZ^{[1]}_{2 \,\text{centre}}$ one-form symmetry (as discussed in \cite[Section 2.4]{Cordova:2017vab}).  Since there is no mixed anomaly between $(\BZ^{[1]}_{2})_{\hat{\CC}}$ and $(\BZ^{[1]}_{2})_{\text{centre}}$ in \eref{SESIL}, there is no non-trivial extension between $ (\BZ^{[0]}_{2})_{\CC}$ and $(\BZ^{[1]}_{2})_{\text{centre}}$ (and so the exact sequence \eref{SESIIL} is split). 
Given that we identified $(\BZ^{[0]}_{2})_{\CC} \equiv (\BZ^{[0]}_{2})_{g''}$, the same statements hold on the side of theory II(R) between $(\BZ^{[0]}_{2})_{g''}$ and $(\BZ^{[1]}_{2})_{\text{centre}}$.
This is consistent with the fact that the description II(R) has a $\BZ_2^{[1]}$ one-form symmetry \cite{Bergman:2020ifi}, without any extension with a zero-form symmetry.  We will describe how to see the relation between $(\BZ^{[0]}_{2})_{\CC} \equiv (\BZ^{[0]}_{2})_{g''}$ and $U(1)^{[0]} \times \BZ^{[0]}_{\GCD(N,2)}$ in terms of the index around \eref{defwpgpp}.

Finally, we gauge the $\BZ_2^{[1]}$ one-form symmetry in each description of the duality II. As a result, we gain a $\BZ_2^{[0]}$ zero-form symmetry in both descriptions of the duality III.  For convenience, let us denote it by $(\BZ_{2}^{[0]})_{\fg}$.  For theory III(L), there is also a magnetic zero-form symmetry, denoted by $(\BZ^{[0]}_{2})_{\zeta}$.\footnote{In the literature, the zero-form magnetic symmetry is usually denoted by $\CM$. Since we use the fugacity $\zeta$ for the magnetic symmetry when we discuss superconformal indices, we denote such a symmetry by $(\BZ^{[0]}_{2})_{\zeta}$ to be consistent with the notation adopted in the subsequent sections of the paper. Notice also that in the special case of $N=1$ this symmetry is actually continuous $U(1)^{[0]}_\zeta$.} By the same reasoning as in \cite[Section 2.4 and Footnote 20]{Cordova:2017vab}, the (abelian) zero-form symmetry of the description III(L) is either $(\BZ_{2}^{[0]})_{\fg} \times (\BZ_{2})_\zeta^{[0]}$ if $2N$ in the $SO(2N)$ gauge factor is not equal to $2$ $\mod$ $4$, or the non-trivial extension $\BZ_{4}^{[0]}$ of the former if $2N$ in the $SO(2N)$ gauge factor is equal to $2$ $\mod$ $4$.\footnote{\label{footN1}The above statements hold only for $N>1$ since eq.~(2.18) of \cite{Cordova:2017vab} for the anomaly applies only to discrete symmetries, while the case $N=1$ where the magnetic symmetry is $U(1)^{[0]}_\zeta$ should be treated separately. In \eqref{ntextIIIL} we show at the level of the index that $(\BZ_{2}^{[0]})_{\fg}$ can be absorbed into $U(1)^{[0]}_\zeta$, indicating that the non-trivial extension between the two symmetries occurs also for $N=1$.}  On the other hand, in theory III(R), there is a non-trivial extension $\BZ^{[0]}_{4}$ zero-form symmetry that arises from the mixed anomaly \eref{SESIIL} between the discrete topological symmetry $(\BZ^{[0]}_{2})_{g''}$ of the theory II(R) and the one-form symmetry $(\BZ^{[1]}_{2})_{\text{centre}}$, which we gauged.  As pointed out in \cite[(2.17)]{Bergman:2020ifi}, the description III(R) has a $U(1)^{[0]} \times \BZ^{[0]}_{\GCD(N,4)}$ zero-form symmetry, which is a further non-trivial extension between the aforementioned $\BZ_4^{[0]}$ zero-form symmetry and the topological symmetry $U(1)^{[0]}_{\text{top}}$:\footnote{For $N=1$ the non-trivial extension implies that the symmetry is only $U(1)^{[0]}$. This is compatible with what happens on the side of theory III(L) as discussed in Footnote \ref{footN1}. We show that $U(1)^{[0]}$ on the side III(R) is identified with $U(1)^{[0]}_\zeta$ on the side III(L) at the level of the index in \eqref{indmatchIIIN1}.}
\bes{ \label{SESIIIR}
\text{III(R)}: \quad 0~\rightarrow~\BZ^{[0]}_{4}~\rightarrow~ U(1)^{[0]} \times \BZ^{[0]}_{\GCD(N,4)}~\rightarrow~U(1)^{[0]}_{\text{top}}~\rightarrow~  0
}
Note that the $U(1)^{[0]}_{\text{top}}$ topological symmetry is not manifest in theory III(L).  As a result, not every generator of the $U(1)^{[0]} \times \BZ^{[0]}_{\GCD(N,4)}$ symmetry is manifest in theory III(L).  For example, in \eref{matchindicesIIIN2}, we show that in the case of $N=2$, the $\BZ^{[0]}_2$ subgroup of the $U(1)^{[0]}$ symmetry of theory III(R) is identified with the  $\BZ^{[0]}_2$ subgroup of the $(\BZ_{2}^{[0]})_{\fg} \times (\BZ_{2}^{[0]})_\zeta$ symmetry or the $\BZ_{4}^{[0]}$ symmetry of theory III(L).

\subsubsection*{Special cases}

There are many interesting special cases that can be considered.  
\ben
\item For $N=1$, duality II becomes
\bes{ \label{U12U1m2duality}
SO(2)_2 \times USp(2)_{-1}  \,\, \leftrightarrow \,\, [U(1)_4 \times U(1)_{-4}]/\BZ_2 \,\, \leftrightarrow \,\, U(1)_2 \times U(1)_{-2}~.
}
We discuss the indices of these theories in Section \ref{sec:dualtiyIIN1}.  They are different descriptions of the worldvolume theory of a single M2-brane on $\BC^4/\BZ_2$ singularity, and so they all have $\CN=8$ supersymmetry.  The second arrow is, in fact, a special case of the following duality for abelian theories:
\bes{ \label{genquotientabelian}
[U(1)_{kp} \times U(1)_{-kp}]/\BZ_p \quad \longleftrightarrow \quad U(1)_k \times U(1)_{-k}~.
}
\item The theories involved in duality III are also related to others as follows. 
\bes{ \label{dualSO2N2USp2Nm1modZ2}
[SO(2N)_2 \times  USp(2N)_{-1}]/\BZ_2   &\quad \longleftrightarrow \quad  [U(N)_4 \times U(N)_{-4}]/\BZ_4 \\
 &\quad \overset{\text{\cite{Honda:2012ik, Tachikawa:2019dvq, Bergman:2020ifi}}}{\longleftrightarrow} \quad  [SU(N)_4 \times SU(N)_{-4}]/\BZ_N
}
\ben
\item For $N=1$, we have the theory of two free hypermultiplets:
\bes{ \label{dualtwofreehyps}
[SO(2)_2 \times  USp(2)_{-1}]/\BZ_2   \quad &\leftrightarrow \quad  [U(1)_4 \times U(1)_{-4}]/\BZ_4 \\
&\leftrightarrow \quad  \text{2 free hypermultiplets} \\
&\overset{\text{\eref{genquotientabelian}}}{\leftrightarrow} \quad  U(1)_1 \times U(1)_{-1}
}
\item For $N=2$, we have
\bes{
[SO(4)_2 \times  USp(4)_{-1}]/\BZ_2   \quad &\leftrightarrow \quad  [U(2)_4 \times U(2)_{-4}]/\BZ_4 \\
&\overset{\text{\cite{Honda:2012ik, Tachikawa:2019dvq, Bergman:2020ifi}}}{\leftrightarrow}  \quad  [SU(2)_4 \times SU(2)_{-4}]/\BZ_2 \\
&\overset{\text{\cite{Tachikawa:2019dvq}}}{\leftrightarrow} \quad  U(3)_2 \times U(2)_{-2} \\
&\overset{\text{\cite{Tachikawa:2019dvq}}}{\leftrightarrow}  \quad  \text{$Spin(5)/\BZ_2$ or $USp(4)/\BZ_2$ SYM}
}
These theories have $\CN=8$ supersymmetry. On the other hand, we find that the theory $SO(4)_2 \times  USp(4)_{-1}$, which is dual to $[U(2)_4 \times U(2)_{-4}]/\BZ_2$,  has $\CN=6$ supersymmetry; see Section \ref{sec:U24U2m4modZ2}.  The $\BZ_2$ discrete quotient, indeed, brings about extra operators carrying a non-trivial charge under the new $\BZ_2$ zero-form topological symmetry. The conserved currents associated to these operators lead to $\CN=8$ supersymmetry.  
\item On the other hand, for the case of $N=3$ in \eref{dualSO2N2USp2Nm1modZ2}, from the index computation, we see that $[U(3)_4 \times U(3)_{-4}]/\BZ_4$ possesses $\CN=6$ supersymmetry; see the comment below \eref{traceM}.
\een
\een

\subsubsection*{Generalisations}
The above results can be generalised in many ways.  First, we consider the $U(3)_4 \times U(1)_{-4}$ and its dual $O(4)_2 \times USp(2)_{-1}$.  According to the discussion around \cite[(3.19)]{Tachikawa:2019dvq}, such theories have a non-anomalous $\BZ_2$ one-form symmetry.  Upon gauging this symmetry, we obtain a duality pair: $[U(3)_4 \times U(1)_{-4}]/\BZ_2 \, \leftrightarrow \, SO(4)_2 \times USp(2)_{-1}$.  We discuss the symmetries of these theories in Sections \ref{sec:SO4USp2} and \ref{sec:O4USp2}.

We then generalise \eref{dualSO2N2USp2Nm1modZ2} to a circular quiver with alternating $SO(2)_2$ and $USp(2)_{-1}$ gauge groups, with a discrete $\BZ_2$ quotient.  It turns out that the theories in this class are dual to 3d $\CN=4$ gauge theories described by a circular quiver with a collection of $U(1)$ gauge groups and with a hypermultiplet with charge 1 under each gauge group; see \eref{KN}.  The detail is provided in Section \ref{sec:circular}.

Finally, we study the dual pair $O(2N+1)_2 \times USp(2)_{-1} \, \leftrightarrow \, U(N+1)_4 \times U(N)_{-4}$, as well as the dual pair $SO(2N+1)_2 \times USp(2)_{-1}  \, \leftrightarrow \,  [U(N+1)_4 \times U(N)_{-4}]/\BZ_2$.  As a surprise, it turns out that these four theories have the same superconformal indices, even refined with fugacities for their 0-form discrete symmetries; see Section \ref{sec:SOodd}. In particular, the zero-form charge conjugation symmetry in the $SO(2N+1)_2 \times USp(2)_{-1}$ theory acts trivially and is unfaithful, so as the $\BZ_2$ zero-form symmetry arising from the $\BZ_2$ discrete gauging in the $[U(N+1)_4 \times U(N)_{-4}]/\BZ_2$ theory.  We conjecture that the $\BZ_2$ one-form symmetry of the first two theories acts trivially on the spectrum of the line operators.

\section{Dualities and superconformal indices} \label{sec:indices}
\subsection{$[SO(2N)_2 \times  USp(2N)_{-1}]/\BZ_2   \,\, \leftrightarrow \,\, [U(N)_4 \times U(N)_{-4}]/\BZ_4$}
In this subsection, we consider the duality between these two theories:
\bes{
\text{III(L):}~ [SO(2N)_2 \times  USp(2N)_{-1}]/\BZ_2   \quad &\leftrightarrow \quad  \text{III(R):}~ [U(N)_4 \times U(N)_{-4}]/\BZ_4
}
\subsubsection{The case of $N=1$} \label{sec:IIIN1}
For $N=1$, the theory III(R): $[U(1)_4 \times U(1)_{-4}]/\BZ_4$ is dual to $SU(1)_4 \times SU(1)_{-4}$ \cite{Tachikawa:2019dvq, Bergman:2020ifi}.  We expect the latter to be identical to the theory of two free hypermultiplets, which is also dual to the $U(1)_1 \times U(1)_{-1}$ theory.  Subsequently, we study these theories in detail with the aid of the superconformal index.

The index of theory III(R) is given by (we summarise our conventions for the index in Appendix \ref{app:index}, see in particular \eqref{indchir} for the contribution $\mathcal{Z}_{\text{chir}}$ of the chiral multiplet)
\bes{ \label{indexIIIR}
\CI^{N=1}_{\text{III(R)}}(u, v, w) &= \sum_{p=0}^3 g^p \sum_{(m_1; m_2) \in \left( \BZ + \frac{p}{4} \right)^2} \oint \frac{d z_1}{2 \pi i z_1} \oint \frac{d z_2}{2 \pi i z_2}  z_1^{4 m_1} z_2^{-4 m_2} w_1^{m_1} w_2^{m_2} \times \\  
& \qquad \prod_{s = \pm 1}  \CZ_{\text{chir}}( u^s z_1 z_2^{-1}; m_1-m_2; 1/2)   \CZ_{\text{chir}}( v^s z_2 z_1^{-1}; m_2-m_1; 1/2)  \\
&=\prod_{s= \pm 1} \CZ_{\text{chir}} \left(g^3 u^{s} \tilde{w}^{-1/2};0 ; 1/2 \right) \CZ_{\text{chir}} \left(g v^{s} \tilde{w}^{1/2};0; 1/2 \right) \\
&=\prod_{s= \pm 1} \CZ_{\text{chir}} \left(w^{-1} u^{s};0 ; 1/2 \right) \CZ_{\text{chir}} \left(w v^{s};0; 1/2 \right) \\
}
where $u$ and $v$ are the fugacities for the $SU(2)_u \times SU(2)_v$ flavour symmetry and $w$ is the fugacity for the $U(1)^{[0]}_w$ zero-form topological symmetry.  In the above, $g$ is the $\BZ_4$ discrete topological fugacity satisfying $g^4=1$, but it can be absorbed into a $U(1)$ global symmetry by a redefinition.  In particular, we have defined 
\bes{ \label{redefw}
\tilde{w} = (w_1 w_2)^{1/2}~, \qquad w = g \tilde{w}^{1/2}=g(w_1 w_2)^{1/4}~.}

We remark that if one makes a change of variables $s_1 =z_1 z_2$ and $s_2 = z_1 z_2^{-1}$, the contribution of the matter fields $\CZ_{\text{chir}}$ is independent of $s_1$ and so the integration over $s_1$ leads to a delta-function that sets (see also \cite[(4.15)]{Tachikawa:2019dvq})
\bes{ \label{fluxpair1}
m_1=m_2~.
}  
This is in agreement with the discussion of \cite[Section 4.1.4]{Cremonesi:2016nbo}. As a result, only the combination $w_1 w_2$, but not $w_1/w_2$, appears in the index. 

The last line of \eref{indexIIIR} is indeed the index of the theory with two free hypermultiplets.  These are identified as the gauge invariant dressed di-baryons, discussed in \cite[(2.11)]{Bergman:2020ifi} (with $N=1$ and $k=4$ in their notation):
\bes{
\CB_{\alpha} = T_{\{-\frac{1}{4}; -\frac{1}{4}\}} A_{\alpha}~, \quad \CB'_{\alpha'} = T_{\{\frac{1}{4}; \frac{1}{4}\}} B_{\alpha'} ~;
}
where $T_{\{m;n\}}$ denotes a monopole operator with fluxes $m$ and $n$ in the first and second $U(1)$ gauge groups respectively.  Its gauge charge is $(k m, -k n)$ under the first and second $U(1)$ gauge groups.  We denote by $\alpha, \beta, \ldots =1, 2$ the indices for the $SU(2)_u$ flavour symmetry and by $\alpha',\beta', \ldots =1,2$ the indices for the $SU(2)_v$ flavour symmetry. Note that the gauge invariant dressed monopole operators
\bes{
(\CM_{-1})_{\alpha_1 \cdots \alpha _4} &= T_{\{-1; -1\}} A_{\alpha_1}\cdots A_{\alpha_4}~, \\
(\CM_{+1})_{\alpha'_1 \cdots \alpha' _4} &= T_{\{+1; +1\}} B_{\alpha'_1}\cdots B_{\alpha'_4}~, \\
}
are related to the di-baryons by the relations
\bes{
(\CM_{-1})_{\alpha_1 \cdots \alpha _4} = \prod_{j=1}^4 \CB_{\alpha_j}~, \quad (\CM_{+1})_{\alpha'_1 \cdots \alpha' _4} = \prod_{j=1}^4 \CB'_{\alpha'_j}~.
}

In order to obtain the index for $U(1)_1 \times U(1)_{-1}$, we proceed as follows. We rewrite the above index using the variables $\tilde{m}_1 = 4m_1$ and $\tilde{m}_2= 4m_2$.  The contribution from the Chern-Simons levels is therefore $z_1^{\tilde{m}_1} z_2^{\tilde{m}_2}$. The summation of $(m_1, m_2) \in (\BZ + p/4)^2$ is then equivalent to the summation of $(\tilde{m}_1, \tilde{m}_2) \in (4\BZ+p)^2$, where $p$ is summed from $0$ to $3$.  
The factors corresponding to the topological fugacities are $w_1^{m_1} w_2^{m_2} = w_1^{\frac{1}{4}\tilde{m}_1} w_2^{\frac{1}{4}\tilde{m}_2}$. We can now shift $\tilde{m}_{1,2} \rightarrow \tilde{m}_{1,2}+p$ and so, together with $g^p$, we have
\bes{ \label{changevars1}
(g (w_1 w_2)^{\frac{1}{4}})^p w_1^{\frac{1}{4}\tilde{m}_1} w_2^{\frac{1}{4}\tilde{m}_2} = w^p w_1^{\frac{1}{4}\tilde{m}_1} w_2^{\frac{1}{4}\tilde{m}_2}
}
where $w=g(w_1 w_2)^{\frac{1}{4}}$ as stated in \eref{redefw}.   Using \eref{fluxpair1}, namely $\tilde{m}_1=\tilde{m}_2 \equiv \tilde{m}$, and writing $w_1 = w s$ and $w_2 =w s^{-1}$, \eref{changevars1} becomes $w^{p+\frac{1}{2} \tilde{m}}$.  Upon shifting $\tilde{m} \rightarrow \tilde{m} - 2p$, we are left with $w^{\frac{1}{2} \tilde{m}}= w^{\frac{1}{4}\tilde{m}_1} w^{\frac{1}{4}\tilde{m}_2}$.  Observe that the discrete fugacity $g$ as well as the factor $w^p$ disappear from the index.  At this point, the summation of $p \in \{0,1,2,3\}$ together with the summation of $(\tilde{m}_1; \tilde{m}_2) \in \left(4\BZ+p\right)^2$ can be replaced by the summation of $(\tilde{m}_1; \tilde{m}_2) \in \BZ^2$.   Moreover, the argument in $\CZ_{\text{chir}}$ depends only on $m_1-m_2$, which is an integer, and so we can replace it by $\tilde{m}_1- \tilde{m}_2$.    Overall, we obtain the index for $U(1)_1 \times U(1)_{-1}$, as required.  This statement can be easily generalised to show duality \eref{genquotientabelian}, namely $[U(1)_{kp} \times U(1)_{-kp}]/\BZ_p \, \leftrightarrow \, U(1)_{k} \times U(1)_{-k}$.

The index of theory III(L): $[SO(2)_2 \times  USp(2)_{-1}]/\BZ_2$ is given by
\bes{ \label{indexIIIL}
&\CI^{N=1}_{\text{III(L)}}(f, \omega) \\
&= \sum_{p=0}^1 \mathfrak{g}^p \sum_{(m_1, m_2) \in \left( \BZ + \frac{p}{2} \right)^2 } \oint \frac{d z_1}{2 \pi i z_1} \oint \frac{d z_2}{2 \pi i z_2}  z_1^{2 m_1} \zeta^{m_1} z_2^{-2 m_2} \CZ^{USp(2)}_{\text{vect}}(z_2, m_2)  \times \\  
& \quad  \prod_{s, s_1, s_2 = \pm 1}  \CZ_{\text{chir}}( f^s z_1^{s_1} z_2^{s_2}; s_1 m_1+{s_2} m_2; 1/2) \\
&= \prod_{s= \pm 1} \CZ_{\text{chir}} \left(\mathfrak{g}\, \zeta^{\frac{1}{2}} f^{s};0 ; 1/2 \right) \CZ_{\text{chir}} \left(\mathfrak{g}\, \zeta^{-\frac{1}{2}} f^{s};0 ; 1/2 \right) \\
&= \prod_{s= \pm 1} \CZ_{\text{chir}} \left(\omega f^{s};0 ; 1/2 \right) \CZ_{\text{chir}} \left(\omega^{-1} f^{s};0 ; 1/2 \right)
}
where $\zeta$ is the fugacity of the $U(1)^{[0]}_\zeta$ magnetic (topological) symmetry of $SO(2)$, $f$ is the fugacity of the $SU(2)_f$ flavour symmetry, $\mathfrak{g}$ is the fugacity associated with the $\BZ_2$ topological symmetry, the fugacity $\omega$ is defined as
\bes{\label{ntextIIIL}
\omega = \mathfrak{g}\, \zeta^{\frac{1}{2}}~,
}
and the contribution $\CZ^{USp(2)}_{\text{vect}}$ from the $USp(2)$ vector multiplet is as defined in \eqref{induspvect}.
Observe that the $\BZ_2$ zero-form symmetry, associated with the fugacity $\mathfrak{g}$, can be absorbed into the magnetic symmetry. This results in the $U(1)^{[0]}_\omega$ zero-form global symmetry.  The last line of \eref{indexIIIL} is indeed the index of the theory with two free hypermultiplets.  These are identified with the gauge invariants dressed di-baryons:
\bes{
\CB^+_\alpha = T_{\left \{ \frac{1}{2}; \frac{1}{2} \right \}} A_\alpha~, \qquad \CB^-_\alpha = T_{\left \{- \frac{1}{2}; -\frac{1}{2} \right \}} A_\alpha~.
}
where in this case $\alpha=1,2$ is an index for the $SU(2)_f$ flavour symmetry.

Comparing \eref{indexIIIR} with \eref{indexIIIL}, we see that 
\bes{\label{indmatchIIIN1}
\CI^{N=1}_{\text{III(L)}}(f, \omega) = \CI^{N=1}_{\text{III(R)}}(u=f, v=f, w=\omega)~.
}
The $U(1)^{[0]}_w$ zero-form symmetry of theory III(R) is mapped to the $U(1)^{[0]}_\omega$ zero-form symmetry of theory III(L), whereas the $SU(2)_f$ flavour symmetry of theory III(L) is identified with the diagonal subgroup of the $SU(2)_u \times SU(2)_v$ flavour symmetry of theory III(R).

\subsubsection{$[U(N)_4 \times U(N)_{-4}]/\BZ_4$ with $N \geq 2$}
The index for theory III(R): $[U(N)_4 \times U(N)_{-4}]/\BZ_4$ can be written as
\bes{ \label{indexIIIRb}
&\CI_{\text{III(R)}}(u, v, w_1, w_2, g) \\
&= \sum_{p=0}^3 g^p \sum_{\CS_p} \left( \prod_{j=1}^N \oint \frac{d z^{(j)}_1}{2 \pi i z^{(j)}_1}  \oint \frac{d z^{(j)}_2}{2 \pi i z^{(j)}_2}  (z^{(j)}_1)^{4  m^{(j)}_1} (z^{(j)}_2)^{-4  m^{(j)}_2} \right)  \times \\
& \qquad  w_1^{\sum_{j=1}^N m^{(j)}_1} w_2^{\sum_{j=1}^N m^{(j)}_2}  \prod_{\ell=1}^2 Z^{U(N)}_{\text{vect}}(z^{(1)}_\ell, \ldots, z^{(N)}_\ell; m^{(1)}_\ell, \ldots, m^{(N)}_\ell) \times \\  
& \qquad \prod_{i,j=1}^N \prod_{s = \pm 1}  \CZ_{\text{chir}}( u^s z^{(i)}_1/z^{(j)}_2; m^{(i)}_1-m^{(j)}_2; 1/2)   \CZ_{\text{chir}}( v^s z^{(i)}_2/z^{(j)}_1; m^{(i)}_2-m^{(j)}_1; 1/2)
}
where $g^4=1$, the notation $\CS_p$ stands for the summation over 
\bes{(m^{(1)}_1, \ldots, m^{(N)}_1 ; m^{(1)}_2, \ldots, m^{(N)}_2) \in \left( \BZ + \frac{p}{4} \right)^{2N}} and the contribution $Z^{U(N)}_{\text{vect}}$ of the $U(N)$ vector multiplet is as defined in \eqref{induspvect}.

As explained in \cite[Section 4.1.4]{Cremonesi:2016nbo}, due to the $D$-term equations, the gauge invariant quantities can be formed provided that the magnetic fluxes of the two gauge groups are paired:\footnote{Note that the integration over the diagonal gauge $U(1)$ leads to a delta-function imposing the constraint $\sum_{j=1}^Nm^{(j)}_1=\sum_{j=1}^Nm^{(j)}_2$.}
\bes{ \label{relationm1m2}
m^{(j)}_1 = m^{(j)}_2 \equiv m^{(j)} ~, \quad j=1, \ldots, N~,
}
provided that we use the Weyl symmetry to order fluxes so that $m^{(1)}_1 \geq m^{(2)}_1 \geq \cdots \geq m^{(N)}_1$ and  $m^{(1)}_2 \geq m^{(2)}_2 \geq \cdots \geq m^{(N)}_2$.  As a result, only the combination of $w_1 w_2$ appears in the index.

According to \cite[Section 2.3]{Tachikawa:2019dvq}, the apparent $\BZ_4$ zero-form symmetry, associated with the fugacity $g$, is actually $\BZ_{\GCD(N,4)}$. More generally, in the $[U(N)_k \times U(N)_{-k}]/\BZ_k$ theory part of the $\mathbb{Z}_k$ zero-form symmetry can be absorbed into the $U(1)$ topological symmetry and only $\BZ_{\GCD(N,k)}$ remains.
By explicitly computing the index, one can indeed check that it can be rewritten solely in terms of\footnote{We normalise the power of $w$ such that the di-baryon operators, which involve the monopole operators $T_{\scriptsize \pm\{\underbrace{1/k,\cdots , 1/k}_N\, ; \,\,\underbrace{1/k, \cdots, 1/k}_{N} \}}$, carry $U(1)^{[0]}_w$ charge $\pm 1$; see \eref{opsorderx}. This explains the power $N/k$ of $w_1 w_2$ in \eref{redef1}. \label{foot:defw}}
\bes{ \label{redef1}
w = g (w_1 w_2)^{N/k}~, \quad g' = g^{k/\GCD(N,k)}~,
}
This means that the actual zero-form global symmetry of the theory is $SU(2)_u \times SU(2)_v \times U(1)^{[0]}_w \times \left( \BZ^{[0]}_{\GCD(N,k)} \right)_{g'}$.

\subsubsection*{The special case of $N=2$}
After redefining the fugacities as in \eref{redef1}
\bes{
w = g(w_1 w_2)^{1/2}~, \quad g' = g^2~,
}
the index \eref{indexIIIRb} for the case of $N=2$ can be written as
\bes{ \label{indexIIIRbN2}
&\CI^{N=2}_{\text{III(R)}}(u, v, w, g') \\
&= 1+x \left[ \chi^{SU(2)}_{[1]} (u)\chi^{SU(2)}_{[1]} (v) + w^{-1} \chi^{SU(2)}_{[2]} (u)+ w \chi^{SU(2)}_{[2]} (v)\right] \\
&\quad +x^2\Big[ (g'+1) w^{-2} \chi^{SU(2)}_{[4]} (u)+(g'+1) w^{-1}\chi^{SU(2)}_{[3]} (u) \chi^{SU(2)}_{[1]} (v)  \\
& \qquad + \left( w \leftrightarrow w^{-1}, u \leftrightarrow v \right) +(g'+2) \chi^{SU(2)}_{[2]} (u) \chi^{SU(2)}_{[2]} (v) + w^2+w^{-2} \\
& \qquad - \chi^{SU(2)}_{[2]} (u)-  \chi^{SU(2)}_{[2]} (v) \Big]+\ldots~.
}
The unrefined index is (cf. \cite[(4.5)]{Gang:2011xp})
\bes{ \label{indexIIIRbN2unref}
\CI^{N=2}_{\text{III(R)}}(u=1, v=1, w=1, g'=1) &=1 + 10 x + 75 x^2 + 230 x^3 + 449 x^4+ \ldots~.
}
Note that the operators with $R$-charge 1 are\footnote{\label{footmonodress} As pointed out in \cite{Kapustin:2006pk, Klebanov:2008vq, Benna:2009xd, Benini:2011cma, Bergman:2020ifi}, for a $U(N)_k$ gauge group, the monopole operators with the magnetic fluxes $(m_1, m_2, \cdots, m_N)$, with $m_1 \geq m_2 \geq \cdots \geq m_N$, transform under the representation of the $SU(N)$ gauge factor with the Dynkin label $[k(m_1-m_2), k(m_2-m_3), \cdots, k(m_{N-1}- m_N)]$ and carry $U(1)$ gauge charge $k \sum_{i=1}^N m_i$. Consequently, the monopoles $T_{\{\pm\frac{1}{4}, \pm\frac{1}{4}; \pm\frac{1}{4}, \pm\frac{1}{4}\}}$ have charge $(\pm2,\mp2)$ under the $U(1)$ subgroups of the two $U(1)\cong SU(2)\times U(1)$ gauge groups. Moreover, in \eqref{opsorderx} the $\epsilon$-tensors are with respect to the $SU(2)$ parts. Hence, the object $(A_{\alpha})^a_i (A_\beta)^b_j \epsilon^{ij} \epsilon_{ab}$ is invariant under the $SU(2)$ parts, while it has charge $(\pm2,\mp2)$ under the $U(1)$ parts, making the operator $\CB_{\alpha \beta}$ gauge invariant and similarly for $\CB'_{\alpha' \beta'}$.}
\bes{ \label{opsorderx}
\chi^{SU(2)}_{[1]} (u)\chi^{SU(2)}_{[1]} (v):  \qquad & M_{\alpha \alpha'} = (A_\alpha)^a_i (B_{\alpha'})^i_a \\
w^{-1} \chi^{SU(2)}_{[2]} (u): \qquad &\CB_{\alpha \beta} = T_{\{-\frac{1}{4}, -\frac{1}{4}; -\frac{1}{4}, -\frac{1}{4}\}} (A_{\alpha})^a_i (A_\beta)^b_j \epsilon^{ij} \epsilon_{ab} \\ 
w \chi^{SU(2)}_{[2]} (v): \qquad & \CB'_{\alpha' \beta'} = T_{\{+\frac{1}{4}, +\frac{1}{4}; +\frac{1}{4}, +\frac{1}{4}\}} (B_{\alpha'})^i_a (B_{\beta'})^j_b  \epsilon_{ij} \epsilon^{ab}
}
where the last two are the gauge invariant dressed di-baryons.  Here $a, b, \ldots =1, 2$ and $i, j, \ldots =1,2$ are the gauge indices for each $U(2)$ gauge group, $\alpha, \beta, \ldots =1, 2$ are the indices for the $SU(2)_u$ flavour symmetry, and $\alpha', \beta', \ldots=1, 2$ are the indices for the $SU(2)_v$ flavour symmetry.  Let us discuss some examples of marginal operators, contributing at order $x^2$ of the index.  The combinations $\CB \CB$ transform in the representation $\Sym^2 [2;0]=[4;0]+[0;0]$ of $SU(2)_u \times SU(2)_v$ and similarly for $\CB' \CB'$: 
\bes{
w^{-2} \chi^{SU(2)}_{[4]} (u) +w^{-2}: \qquad& \CB_{\alpha \beta} \CB_{\gamma \delta}\\
w^2 \chi^{SU(2)}_{[4]} (v) +w^{2}: \qquad& \CB'_{\alpha' \beta'} \CB'_{\gamma' \delta'}~.
}
Note that the index \eref{indexIIIRbN2} can be rewritten in terms of characters of $SU(4)$ representations as follows:
\bes{ \label{indexIIIRbN2SU4}
\eref{indexIIIRb}_{N=2} &=  1+x \chi^{SU(4)}_{[0,0,2]}(\vec s) \\
& \quad +x^2 \Big[ (g'+1)\chi^{SU(4)}_{[0,0,4]}(\vec s)+ \chi^{SU(4)}_{[0,2,0]}(\vec s) -\chi^{SU(4)}_{[1,0,1]}(\vec s)   \Big]+\ldots~,
}
where we have taken 
\bes{
w= q^2
}
and have used the fugacity map\footnote{In this convention, the character of the fundamental representation $[1,0,0]$ of $SU(4)$ is written as $\chi^{SU(4)}_{[1,0,0]}(\vec s)= s_1 + s_2 s_1^{-1} + s_3 s_2^{-1} + s_3^{-1}$.}
\bes{ \label{fugmap0}
s_1 = q u~, \quad s_2 =q^2~, \quad s_3 = q v~.
}

Since this theory is known to be dual to the Bagger-Lambert-Gustavson \cite{Bagger:2007jr, Gustavsson:2007vu} theory $[SU(2)_4 \times SU(2)_{-4}]/\BZ_2$ and the $USp(4)/\BZ_2$ super-Yang-Mills (see \cite[Table 1]{Tachikawa:2019dvq}), it has $\CN=8$ supersymmetry.  This can be seen from the index \eref{indexIIIRb} as follows. (The argument given below is the same as that of \cite[Appendix C.1]{Beratto:2020qyk}.\footnote{See also \cite{Garozzo:2019ejm,Gang:2018huc,Gang:2021hrd} for other examples of supersymmetry enhancements in 3d detected with the index.}) We rewrite \eref{indexIIIRbN2}, or equivalently \eref{indexIIIRbN2SU4}, as an $\CN=3$ index.\footnote{As a requirement of an $\CN=3$ index, the order $x$ receives a contribution solely from the $\CN=3$ flavour current, and so the coefficient of $x$ must be an adjoint representation of the flavour symmetry of the corresponding $\CN=3$ theory.  The index \eref{indexIIIRbN2}, or equivalently \eref{indexIIIRbN2SU4}, is an $\CN=2$ index, not an $\CN=3$ index, since the coefficient of order $x$ is not an adjoint representation of $SU(4)$. As can be seen below, one can rewrite this as an $\CN=3$ index by tuning some fugacities to be equal and reexpressing the index in terms of characters of representations of $USp(4)$.}  This can be achieved by setting $u=v=f$ and $w=q^2$ in \eref{indexIIIRbN2}, or by using branching rules of representations of $SU(4)$ to those of a maximal subgroup of $USp(4)$ in \eref{indexIIIRbN2SU4}. Either way, we obtain the $\CN=3$ index in terms of characters of representations of $USp(4)$ as
\bes{ \label{indexIIIRbN2a}
\eref{indexIIIRb}_{N=2}
&=  1+x\chi^{USp(4)}_{[2,0]}( \vec h ) + x^2\Big[(g'+1)\chi^{USp(4)}_{[4,0]}( \vec h ) + \chi^{USp(4)}_{[0,2]}( \vec h )+1 \\
& \qquad \qquad +{\blue  \chi^{USp(4)}_{[0,1]}( \vec h )}  - \chi^{USp(4)}_{[2,0]}( \vec h )- {\red \chi^{USp(4)}_{[0,1]}( \vec h )} \Big] +\ldots
}
where we have used the fugacity map\footnote{In this convention, the character of the fundamental representation $[1,0]$ of $USp(4)$ is written as $\chi^{USp(4)}_{[1,0]}( \vec h ) = h_1 + h_1^{-1} + h_2 + h_2^{-1}$.}
\bes{ \label{fugmap1}
&h_1 =  q f ~, \quad h_2 = q^{-1} f \\ 
&s_1 = h_1~, \quad s_2= h_1 h_2^{-1}~, \quad s_3= h_1~.
}
Note that \eref{indexIIIRbN2a} satisfies all of the necessary conditions for the enhanced $\CN=8$ supersymmetry \cite{Evtikhiev:2017heo}.  The blue term in \eref{indexIIIRbN2a} is the contribution of 5 marginal operators in the representation $[0,1]$ of $USp(4)$, whereas the red term in \eref{indexIIIRbN2a} is the contribution of the extra supersymmetry currents. These two contributions precisely cancel with each other.  Since we have 5 extra supersymmetry currents, supersymmetry gets enhanced from $\CN=3$ to $\CN=3+5=8$, as expected.  Let us discuss the marginal operators corresponding to the blue term in \eref{indexIIIRbN2a} in detail. First of all, since $[0,1]$ is a subrepresentation of $\Sym^2[2,0] = [4,0] \oplus [0,2] \oplus [0,1] \oplus [0,0]$, we expect that such marginal operators can be constructed by appropriately multiplying those in \eref{opsorderx}.   Secondly, since the representation $[0,1]$ of $USp(4)$ decomposes into those of $SU(2)_f \times U(1)_q$ as $[2]_0 \oplus [0]_{-2} \oplus [0]_{2}$, we propose that the corresponding operators are respectively\footnote{There are other two operators in the representations $[0]_{\mp 2}$ of $SU(2)_f \times U(1)_q$ that are contained in the branching rule of the representation $[4,0]$ of $USp(4)$, namely
\bes{ \nn
&\epsilon^{\alpha \beta} \epsilon^{\gamma \delta}   (A_\alpha)^a_i (A_\beta)^b_j  (B_\gamma)^i_b (A_\delta)^c_k \epsilon^{jk} \epsilon_{ac} T_{\{-\frac{1}{4}, -\frac{1}{4}; -\frac{1}{4}, -\frac{1}{4}\}}~, \\
&\epsilon^{\alpha \beta} \epsilon^{\gamma \delta} (B_\alpha)_a^i (B_\beta)_b^j  (A_\gamma)_i^b (B_\delta)_c^k \epsilon_{jk} \epsilon^{ac} T_{\{+\frac{1}{4}, +\frac{1}{4}; +\frac{1}{4}, +\frac{1}{4}\}}~.
}}
\bes{
(\Tr \, M)\hat{M}_{\alpha \beta}~, \qquad \hat{M}_{\alpha \beta}\CB_{\gamma \delta} \epsilon^{\beta \gamma} \epsilon^{\alpha \delta}~, \qquad \hat{M}_{\alpha \beta}\CB'_{\gamma \delta} \epsilon^{\beta \gamma} \epsilon^{\alpha \delta}~,
}
where $\alpha, \beta, \gamma, \delta=1,2$ are the indices for $SU(2)_f$, which is a diagonal subgroup of $SU(2)_u \times SU(2)_v$, and we have defined
\bes{ \label{traceM}
\Tr\,M = M_{\alpha \beta}\epsilon^{\alpha \beta}~, \qquad \hat{M}_{\alpha \beta} = M_{\alpha \beta}-\frac{1}{2} (\Tr M)\epsilon_{\alpha \beta}~.
}

Finally, let us remark that in the case of $N=2$ the di-baryon operators $\CB$ and $\CB'$ have $R$-charge $1$ and so they contribute at order $x$ of the index, which makes it fulfil the condition for having $\CN=8$ supersymmetry.  For a general $N$, the di-baryons have $R$-charge $N/2$, and so for $N \geq 3$ they contribute at a higher order of the index.  In the latter case, the only contribution at order $x$ comes from the operators $M$, which have 4 components.  We thus expect the theory with $N\geq 3$ to have $\CN=6$ supersymmetry \cite{Evtikhiev:2017heo}.

\subsubsection{$[SO(2N)_2 \times  USp(2N)_{-1}]/\BZ_2$ with $N \geq 2$}
The index for the theory III(L): $[SO(2N)_{2} \times USp(2N)_{-1}]/\BZ_2$ is 
\bes{ \label{indexIIIL}
&\CI_{\text{III(L)}}(f, \mathfrak{g}, \zeta) \\
&= \sum_{p=0}^1 \mathfrak{g}^p \sum_{\CS'_p} \left( \prod_{j=1}^N \oint \frac{d z^{(j)}_1}{2 \pi i z^{(j)}_1} \oint \frac{d z^{(j)}_2}{2 \pi i z^{(j)}_2}  (z^{(j)}_1)^{2 m^{(j)}_1}  (z^{(j)}_2)^{-2 m^{(j)}_1} \right)   \zeta^{\sum_{j=1}^N  m^{(j)}_1}  \times \\
& \quad  \CZ^{SO(2N)}_{\text{vect}}(z^{(1)}_1,\ldots,z^{(N)}_1; m^{(1)}_1, \ldots,m^{(N)}_1)  \CZ^{USp(2N)}_{\text{vect}}(z^{(1)}_2, \ldots,z^{(N)}_2; m^{(1)}_2, \ldots,m^{(N)}_2)  \times \\  
& \quad  \prod_{i,j=1}^N \,  \prod_{s, s_1, s_2 = \pm 1}  \CZ_{\text{chir}} \left( f^s (z^{(i)}_1)^{s_1} (z^{(j)}_2)^{s_2}; s_1 m^{(i)}_1+{s_2} m^{(j)}_2; 1/2 \right) 
}
where $\CS'_p$ stands for the summation over $(m^{(1)}_1, \ldots, m^{(N)}_1 , m^{(1)}_2, \ldots, m^{(N)}_2) \in \left( \BZ + \frac{p}{2} \right)^{2N}$, $\zeta$ is the fugacity associated with the magnetic symmetry of the $SO(2N)$ gauge group satisfying $\zeta^2=1$, and $\fg$ is the fugacity for the topological $\BZ_2$ symmetry satisfying $\fg^2=1$.

Let us provide an explicit expression for $N=2$ up to order $x^2$:
\bes{ \label{indexIIILbN2}
&\CI^{N=2}_{\text{III(L)}}(f, \mathfrak{g}, \zeta) \\
&= 1+ x \left[ 1+(\fg + \zeta + \fg \zeta) \chi^{SU(2)}_{[2]}(f)  \right] + x^2  \Big[5 \chi^{SU(2)}_{[4]}(f) +5 \\
& \qquad \qquad \quad + (\fg + \zeta + \fg \zeta) \left(2\chi^{SU(2)}_{[4]}(f) +2\chi^{SU(2)}_{[2]}(f) \right) -   \chi^{SU(2)}_{[2]}(f) \Big]+\ldots
}
with the unrefined index $\CI^{N=2}_{\text{III(L)}}(f=1, \mathfrak{g}=1, \zeta=1)$ given by \eref{indexIIIRbN2unref}. Note that it is not possible to absorb $\fg$ with a redefinition of $\zeta$ as in \eref{redef1}.  The manifest zero-form global symmetry of this theory is therefore $SU(2)_f \times (\BZ^{[0]}_2)_\fg \times (\BZ^{[0]}_2)_\zeta$.   We can match the indices \eref{indexIIIRbN2} and \eref{indexIIILbN2} as follows:
\bes{\label{matchindicesIIIN2}
\CI^{N=2}_{\text{III(L)}}(f, \mathfrak{g} = \xi, \zeta = \xi) = \CI^{N=2}_{\text{III(R)}}(u=f,v=f, w=\xi, g'=1) \Big |_{\xi^2=1}~.
}
In the theory III(L) only the diagonal subgroup $SU(2)_f$ of the flavour symmetry $SU(2)_u \times SU(2)_v$ of the theory III(R) is manifest. Moreover, the $U(1)^{[0]}_w$ zero-form symmetry of the theory III(R) is not manifest in the theory III(L), but its $\BZ_2$ subgroup is identified with the diagonal $\BZ_2$ symmetry of $(\BZ^{[0]}_2)_\fg \times (\BZ^{[0]}_2)_\zeta$ in III(L).  Furthermore, the $(\BZ^{[0]}_2)_{g'}$ symmetry of the theory III(R) is not manifest in the theory III(L).  Since we claim that the theories III(L) and III(R) are dual to each other, the theory III(L) is expected to have an emergent zero-form symmetry $SU(2)_u \times SU(2)_v \times U(1)^{[0]}_w \times (\BZ^{[0]}_2)_{g'}$.

\subsection{$SO(2N)_2 \times  USp(2N)_{-1} \,\, \leftrightarrow \,\,[U(N)_4 \times U(N)_{-4}]/\BZ_2$}
In this subsection, we consider the duality between the following two theories:
\bes{
\text{II(L):}~ SO(2N)_2 \times  USp(2N)_{-1}   \quad &\leftrightarrow \quad  \text{II(R):}~ [U(N)_4 \times U(N)_{-4}]/\BZ_2~.
}

The theory II(L) can be obtained by gauging the $(\BZ^{[0]}_2)_\fg$ zero-form symmetry of the theory III(L), where at the level of the index this corresponds to summing over $\fg \in \{\pm 1\}$ in \eref{indexIIIL}.  Note also that in the description II(L) there is also a zero-form $(\BZ^{[0]}_{2})_{\CC}$ charge conjugation symmetry, whose fugacity will be denoted by $\chi$.

On the other hand, we can obtain the theory II(R) from the theory III(R) by gauging a $\BZ_2$ subgroup of the zero-form symmetry $U(1)^{[0]}_w \times (\BZ^{[0]}_{\GCD(N,4)})_{g'}$.  From the perspective of the index, this can be done as follows.  First, we rewrite the index \eref{indexIIIRb} using the variables $w$ and $g'$ as indicated in \eref{redef1}.  For convenience, we represent the theory II(R) as $[U(N)_k \times U(N)_{-k}]/\BZ_{\fm'}$, with $k=\fm' \fm=4$ and $\fm'=\fm=2$.  Taking $w = \tilde{g} (w_1 w_2)^{N/k}$ and summing over $\tilde{g} \in \BZ_{\fm'} =\{ \exp(2\pi i \, j/\fm') | \, j =0,1,\ldots, \fm'-1\}$, we are left with the fugacity $(w_1 w_2)^{N/k}$ and $g'$ such that $(g')^{\GCD(N,k)} =1$. By computing the index one can check that we can further redefine\footnote{Similarly to Footnote \ref{foot:defw}, we normalise the power of $w'$ such that the di-baryon operators, which involve the monopole operators $T_{\scriptsize \pm\{\underbrace{1/\fm',\cdots , 1/\fm'}_N\, ; \,\,\underbrace{1/\fm', \cdots, 1/\fm'}_{N} \}}$, carry $U(1)^{[0]}_w$ charge $\pm 1$; see \eref{dibaryonsmprime}. This explains the power $N/\fm' =\fm N/k$ of $w_1 w_2$ in \eref{defwpgpp}.}
\bes{ \label{defwpgpp}
w' = g' (w_1 w_2)^{N/\fm'} =g' (w_1 w_2)^{\fm N/k}~, \qquad g''= (g')^{\frac{\GCD(N,k)}{\GCD(N,\fm')}}~,
}
where 
\bes{
(g'')^{\GCD(N,\fm')} =  (g')^{\GCD( N, k)} = 1~.  
}
The zero-form symmetry of the theory II(R) is therefore $SU(2)_u \times SU(2)_v \times U(1)^{[0]}_{w'} \times (\BZ^{[0]}_{\GCD(N,\fm')})_{g''}$, in agreement with the discussion in \cite[Section 2.5]{Bergman:2020ifi}.  The above discussion can, in fact, be generalised to any $k$, $\fm$ and $\fm'$.

As a result of gauging a $\BZ^{[0]}_2$ zero-form symmetry, both of II(L) and II(R) have a $\BZ^{[1]}_2$ one-form symmetry.

\subsubsection{The case of $N=1$} \label{sec:dualtiyIIN1}
Let us examine the theory II(R): $[U(1)_4 \times U(1)_{-4}]/\BZ_2$.  The index of this theory is almost the same as \eref{indexIIIR}, with two exceptions: the summation over $(m_1, m_2)$ is in $(\BZ + p/2)^2$, and the summation over $p$ is from $p=0$ to $1$.  The index, up to order $x^2$, can be written as
\bes{ \label{indexIIRa}
\CI^{N=1}_{\text{II(R)}} 
&= 1+x \left[ \chi^{SU(2)}_{[1]} (u)\chi^{SU(2)}_{[1]} (v) + w^{-1} \chi^{SU(2)}_{[2]} (u)+ w \chi^{SU(2)}_{[2]} (v)\right] \\
&\quad +x^2\Big[ w^{-2} \chi^{SU(2)}_{[4]} (u)+w^{-1}\chi^{SU(2)}_{[3]} (u) \chi^{SU(2)}_{[1]} (v) + \left( w \leftrightarrow w^{-1}, u \leftrightarrow v \right) \\
& \qquad +\chi^{SU(2)}_{[2]} (u) \chi^{SU(2)}_{[2]} (v) - (w+w^{-1}) \chi^{SU(2)}_{[1]} (u) \chi^{SU(2)}_{[1]} (v) \\
& \qquad - \chi^{SU(2)}_{[2]} (u)-  \chi^{SU(2)}_{[2]} (v) -2\Big]+\ldots~.
}
It turns out that the theory II(R) coincides with the ABJM theory $U(1)_2 \times U(1)_{-2}$.  To see this equivalence, we make the following change of variables: $m_1' = 2m_1$ and $m_2'=2m_2$.  The contribution from the Chern-Simons levels is therefore $z_1^{2m'_1} z_2^{2m'_2}$.  The summation of $(m_1, m_2) \in (\BZ + p/2)^2$ is then equivalent to the summation of $(m'_1, m'_2) \in (2\BZ + p)^2$, where $p$ is summed over $\{ 0, 1 \}$. At this point, we can just set $g=1$ since $\GCD(N,\mathfrak{m})=\GCD(1,2)=1$ and take the summation of $(m'_1, m'_2)$ to be over $\BZ^2$.  Since $\CZ_{\text{chir}}$ only depends on $m_1-m_2$, which is an integer, we can replace the latter by $m'_1-m'_2$.  Overall, we obtain the index of $U(1)_2 \times U(1)_{-2}$.  Due to this equivalence, we conclude that the theory $[U(1)_4 \times U(1)_4]/\BZ_2$ also has $\CN=8$ supersymmetry.

Let us comments on the operators that contribute to the index \eref{indexIIRa}.  From the perspective of the $[U(1)_4 \times U(1)_4]/\BZ_2$ theory, there are monopole operators
\bes{
T_{\pm \frac{m}{2}} \equiv T_{ \pm m  \left\{\frac{1}{2}; \frac{1}{2}\right\}}~, \quad m \in \BZ~,
} 
which carry gauge charges $\pm m (2,-2)$ under the gauge group $U(1)_4 \times U(1)_{-4}$, due to the discrete $\BZ_2$ quotient.  On the other hand, the monopole operators $T_{\pm \frac{1}{2}}$ do not exist in the $U(1)_2 \times U(1)_{-2}$ theory, but there are instead the monopole operators
\bes{
V_{\pm m} \equiv V_{ \pm m \left\{ 1; 1\right\}}~, \quad m \in \BZ~,
}
which carry gauge charges $\pm m(2,-2)$ under the gauge group $U(1)_2 \times U(1)_{-2}$.
The operators that contribute at order $x$ are 
\bes{
\chi^{SU(2)}_{[1]} (u)\chi^{SU(2)}_{[1]} (v): &\qquad M_{\alpha \alpha'} = A_{\alpha} B_{\alpha'} \\
w^{-1} \chi^{SU(2)}_{[2]} (u): &\qquad T_{- \frac{1}{2}} A_{\alpha}  A_{\beta} \,\, \leftrightarrow \,\, V_{-1}  A_{\alpha}  A_{\beta} \\
w \chi^{SU(2)}_{[2]} (v): &\qquad T_{+ \frac{1}{2}} B_{\alpha'}  B_{\beta'} \,\, \leftrightarrow \,\, V_{+1}  B_{\alpha'}  B_{\beta'}
}
Observe that the di-baryon operators in $[U(1)_4 \times U(1)_4]/\BZ_2$ get mapped to the dressed monopole operators in $U(1)_2 \times U(1)_{-2}$.  The marginal operators, contributing to order $x^2$, are
\bes{
w^{-2} \chi^{SU(2)}_{[4]} (u): &\qquad V_{-2} \, A_{\alpha_1} A_{\alpha_2} A_{\alpha_3}  A_{\alpha_4} \\
w^{-1}\chi^{SU(2)}_{[3]} (u) \chi^{SU(2)}_{[1]} (v): &\qquad V_{-1} \, A_{\alpha_1} A_{\alpha_2} A_{\alpha_3} B_{\alpha'_1} \\
\chi^{SU(2)}_{[2]} (u) \chi^{SU(2)}_{[2]} (v):  &\qquad A_{\alpha_1} A_{\alpha_2} B_{\alpha'_1} B_{\alpha'_2}~,
}
where in the first two lines we can obtain those correspond to the terms $w^{2} \chi^{SU(2)}_{[4]} (v)$ and $w\chi^{SU(2)}_{[3]} (v) \chi^{SU(2)}_{[1]} (u)$ by simply simultaneously exchanging $V_{-m} \leftrightarrow V_{+m}$ and $A \leftrightarrow B$.  These gauge invariant combinations are written from the perspective of the $U(1)_2 \times U(1)_{-2}$ theory.  In the $[U(1)_4 \times U(1)_4]/\BZ_2$ duality frame, one simply needs to replace $V_{\pm m}$ by $T_{\pm m/2}$ in the above expressions.

Similarly to \eref{indexIIIRbN2SU4} and \eref{indexIIIRbN2a}, the index of the theory $[U(1)_4 \times U(1)_4]/\BZ_2 \, \cong \, U(1)_2 \times U(1)_{-2}$ can be written in terms of $SU(4)$ characters and $USp(4)$ characters as follows:
\bes{\label{indexIIRb}
\CI^{N=1}_{\text{II(R)}} 
&=  1+x \chi^{SU(4)}_{[0,0,2]}(\vec s) +x^2 \Big[\chi^{SU(4)}_{[0,0,4]}(\vec s)-\chi^{SU(4)}_{[1,0,1]}(\vec s) -1  \Big]+\ldots \\
&= 1+x\chi^{USp(4)}_{[2,0]}( \vec h ) + x^2\Big[\chi^{USp(4)}_{[4,0]}( \vec h )   - \chi^{USp(4)}_{[2,0]}( \vec h )- {\red \chi^{USp(4)}_{[0,1]}( \vec h )}-1 \Big] 
}
where we use the fugacity maps \eref{fugmap0} and \eref{fugmap1}.  The first line should be regarded as an $\CN=2$ index, whereas the second line should be regarded as an $\CN=3$ index, since \eg~ the coefficient of $x$ is an adjoint representation of the flavour symmetry of the $\CN=3$ theory.  The red term is the contribution of the $\CN=3$ extra supersymmetry currents.  Since there are 5 of them in the representation $[0,1]$ of $USp(4)$, supersymmetry gets enhanced from $\CN=3$ to $\CN=3+5=8$.  The term $-1$ at order $x^2$ worths some explanations.  This corresponds to a conserved current associated with the $U(1)$ global symmetry that gives charge 1 to all of the chiral multiplets $A_\alpha$ and $B_{\alpha'}$.  For convenience, we shall denote this symmetry by $U(1)_D$. Note that this symmetry is specific to the abelian ABJM theory, since the superpotential vanishes.  In the non-abelian case, the superpotential $\epsilon^{\alpha \beta} \epsilon^{\alpha' \beta'} A_\alpha A_\beta B_{\alpha'} B_{\beta'}$ does not vanish and so the $U(1)_D$ symmetry is explicitly broken.  Moreover, the current of the $U(1)_D$ symmetry does not belong to the $\CN=3$ flavour current multiplet and so does not contribute at order $x$ of the index. This is because the $\CN=3$ superpotential $\sum_{j=1}^2 (A_j \phi_1 B_j - A_j \phi_2 B_j) + \frac{4}{4\pi} (\phi_1^2-\phi_2^2)$, where $\phi_1$ and $\phi_2$ are complex scalars in the vector multiplets of each gauge group, of the abelian ABJM theory does not allow such a charge assignment.

Let us now analyse the index of the $SO(2)_2 \times USp(2)_{-1}$ theory.  This can be computed using \eref{indexIIIL} with two modifications: the summation of $(m_1, m_2)$ is over $\BZ^2$, and the part $\sum_{p=0}^1 \fg^p$ is removed.  The result is the same as \eref{indexIIRa}, with $u=v=f$ and $w=\zeta$.  In other words, the $SU(2)_f$ flavour symmetry of theory II(L) is identified with the diagonal subgroup of $SU(2)_u \times SU(2)_v$ of theory II(R), and the $U(1)$ magnetic symmetry of theory II(L) is identified with the topological symmetry of theory II(R).  In fact, we can also turn on the fugacity $\chi$ for the zero-form $\BZ^{[0]}_2$ charge conjugation symmetry \cite{Aharony:2013kma}.  For $\chi=1$, the index is the same as that of $SO(2)_2 \times USp(2)_{-1}$, \ie~ as discussed before.  For $\chi=-1$, the index is
\bes{
&\sum_{m_2 \in \BZ} \oint \frac{d z_2}{2 \pi i z_2} z_2^{-2 m_2} \CZ^{USp(2)}_{\text{vect}}(z_2, m_2)  \prod_{s, s_1, s_2 = \pm 1}  \CZ_{\text{chir}}( s_1 f^s z_2^{s_2};{s_2} m_2; 1/2) \\
& = 1+\left(-f^2-\frac{1}{f^2}\right) x+\left(f^4+\frac{1}{f^4}+1\right) x^2+\left(-f^6-\frac{1}{f^6}\right) x^3+\ldots~.
}
Let $\zeta'$ be the fugacity for a $\BZ_2$ subgroup of the $U(1)$ magnetic symmetry.  The index can be written in terms of the fugacities $f$, $\zeta'$ and $\chi$ as
\bes{ \label{indIILN1zc}
&1+ x \left[ 1 +(\zeta' +\chi+\zeta' \chi) \chi^{SU(2)}_{[2]}(f) \right] \\
& \qquad + x^2 \left[ (\zeta' +\chi+\zeta' \chi+2) \chi^{SU(2)}_{[4]}(f)- \chi^{SU(2)}_{[2]}(f) - (\zeta' +\chi+\zeta' \chi) \right]+\ldots
}
where $\zeta'^2=\chi^2=1$. Note that $\zeta'$ and $\chi$ appear on an equal footing and they can be interchanged. In this notation, we can match the indices \eref{indexIIRa} and \eref{indIILN1zc} as
\bes{
[\eref{indexIIRa}](u=f, v=f, w= \chi) \Big|_{\chi^2=1}= [\eref{indIILN1zc}](f, \zeta'=\chi, \chi)~.
}
The unrefined indices of theories II(L) and II(R) with $N=1$ are given by \cite[(4.2)]{Cheon:2012be}:
\bes{
1+10x+19x^2 +26x^3 +49x^4 +26x^5 + \ldots~.
}

\subsubsection{$[U(2)_4 \times U(2)_4]/\BZ_2$} \label{sec:U24U2m4modZ2}
The index of theory II(R): $[U(2)_4 \times U(2)_4]/\BZ_2$ can be written as 
\bes{ \label{indIIR}
&1+x \left[ \chi^{SU(2)}_{[1]} (u)\chi^{SU(2)}_{[1]} (v) \right]  +x^2\Big[ (g''+1) w'^{-1} \chi^{SU(2)}_{[4]} (u)+ (g''+1) w' \chi^{SU(2)}_{[4]} (v)  \\
& \quad  + g'' (w'+w'^{-1}) +(g''+2) \chi^{SU(2)}_{[2]} (u) \chi^{SU(2)}_{[2]} (v) +1  \\
& \quad - \chi^{SU(2)}_{[2]} (u)-  \chi^{SU(2)}_{[2]} (v)-1 \Big]+\ldots
}
where $(g'')^2=1$ and we have used the notation as discussed around \eref{defwpgpp}.  The corresponding unrefined index is given by
\bes{ \label{indIIRunref}
1 + 4 x + 43 x^2 + 108 x^3 + 241 x^4+ \ldots~.
}

The operators that contribute at order $x$ are 
\bes{
\chi^{SU(2)}_{[1]} (u)\chi^{SU(2)}_{[1]} (v): \qquad M_{\alpha \alpha'} = (A_{\alpha})^a_i (B_{\alpha'})^i_a~.
}
Due to the $\BZ_2$ quotient, the elementary monopole operators are 
\bes{
T_{\pm \frac{1}{2}} \equiv T_{  \left\{\pm\frac{1}{2}, \pm \frac{1}{2}; \pm \frac{1}{2}, \pm \frac{1}{2} \right\}}~.
} 
Arising from the discrete gauging, they transform non-trivially under the $(\BZ^{[0]}_2)_{g''}$ zero-form symmetry.  They carry charges $\pm 1$ under the $U(1)^{[0]}_{w'}$ zero-form topological symmetry.  Moreover, $T_{ - \frac{1}{2} }$ carries gauge charges $4\left(-\frac{1}{2}-\frac{1}{2}, \frac{1}{2}+\frac{1}{2} \right) = (-4,+4)$ under the $U(1) \times U(1)$ gauge subgroup of the $U(2)_4 \times U(2)_{-4}$ gauge group. Similarly, $T_{ + \frac{1}{2} }$ carries such gauge charges $(+4, -4)$.

Now let us discuss the marginal operators, contributing the positive terms at order $x^2$.  The di-baryon gauge invariant operators are\footnote{The dressing of the monopole operators works similarly to to what was explained in Footnote \ref{footmonodress}.}
\bes{ \label{dibaryonsmprime}
\scalebox{0.9}{$
\begin{split}
g'' w'^{-1} \left(\chi^{SU(2)}_{[4]} (u)+1 \right): &\quad \CB_{\alpha_1 \ldots \alpha_4} = T_{ -\frac{1}{2}} (A_{\alpha_1})^{a_1}_{i_1} (A_{\alpha_2})^{a_2}_{i_2} (A_{\alpha_3})^{a_3}_{i_3} (A_{\alpha_4})^{a_4}_{i_4} \epsilon^{i_1 i_2} \epsilon_{a_1 a_2} \epsilon^{i_3 i_4} \epsilon_{a_3 a_4} \\
g'' w' \Big(\chi^{SU(2)}_{[4]} (v)+1 \Big): &\quad \CB'_{\alpha'_1 \ldots \alpha'_4} = T_{ +\frac{1}{2} } (B_{\alpha'_1})_{a_1}^{i_1} (B_{\alpha'_2})_{a_2}^{i_2}  (B_{\alpha'_3})_{a_3}^{i_3} (B_{\alpha'_4})_{a_4}^{i_4} \epsilon_{i_1 i_2} \epsilon^{a_1 a_2}\epsilon_{i_3 i_4} \epsilon^{a_3 a_4}
\end{split}$}
}
where the representations $[4] \oplus [0]$ come from the decomposition of $\Sym^2 [2]$ of $SU(2)_u$ or $SU(2)_v$. There are also the following marginal operators:
\bes{
&g'' \chi^{SU(2)}_{[2]} (u) \chi^{SU(2)}_{[2]} (v): \\
&\quad \CG_{\alpha \beta \alpha' \beta' } = T_{\{ \frac{1}{2}, -\frac{1}{2}; \frac{1}{2},- \frac{1}{2} \}} (A_{\alpha})^{a_1}_{i_1} (A_{\beta})^{a_2}_{i_2} (B_{\alpha'})_{a_3}^{i_3} (B_{\beta'})_{a_4}^{i_4} \epsilon^{i_1 i_2} \epsilon_{a_1 a_2} \epsilon_{i_3 i_4} \epsilon^{a_3 a_4}~.
}
There are gauge invariant dressed monopole operators, contributing $w'^{-1} \chi^{SU(2)}_{[4]} (u)$ and $w'\chi^{SU(2)}_{[4]} (v)$ at order $x^2$,
\bes{
(\CM_{-1})_{\alpha_1 \ldots \alpha_4} = (T_{\{-1,0;-1,0\}})_{(a_1\cdots a_4)}^{(i_1 \cdots i_4)} (A_{\alpha_1})^{a_1}_{i_1} (A_{\alpha_2})^{a_2}_{i_2}(A_{\alpha_3})^{a_3}_{i_3}(A_{\alpha_4})^{a_4}_{i_4}  \\
(\CM_{+1})_{\alpha'_1 \ldots \alpha'_4} = (T_{\{+1,0;+1,0\}})^{(a_1\cdots a_4)}_{(i_1 \cdots i_4)} (B_{\alpha'_1})_{a_1}^{i_1} (B_{\alpha'_2})_{a_2}^{i_2}(B_{\alpha'_3})_{a_3}^{i_3}(B_{\alpha'_4})_{a_4}^{i_4}
}
where, as for the ABJM theory, $T_{\pm \{1,0;1,0\}}$ transform in the representation $[4_{\pm 4}; 4_{\mp 4}]$ of the $U(2)_4 \times U(2)_{-4}$ gauge group.\footnote{Here $[k_q]$ stands for a $U(2)\cong SU(2)\times U(1)$ representation consisting of the spin $k/2$ representation of the $SU(2)$ part and having charge $q$ under the $U(1)$ part.}
Finally, there are also the following marginal operators:
\bes{
\chi^{SU(2)}_{[2]} (u) \chi^{SU(2)}_{[2]} (v)+1: \quad  & M_{\alpha \alpha'}M_{\beta \beta'}~, \\
\chi^{SU(2)}_{[2]} (u) \chi^{SU(2)}_{[2]} (v): \quad &\CQ_{\alpha \beta \alpha' \beta'} =(A_\alpha)^a_i (B_{\alpha'})^i_b (A_\beta)^b_j (B_{\beta'})^j_a
}
where the latter are subject to the relations \eref{relationQ} coming from the $F$-terms.  We will discuss these two operators in more detail around \eref{marginalIR}.

Theory II(R): $[U(2)_4 \times U(2)_4]/\BZ_2$, in fact, has $\CN=6$ supersymmetry. This can be seen from the index as follows.  It is convenient to rewrite \eref{indIIR} in terms of an $\CN=3$ index simply by setting $u=v=f$ and using the fact that $[2] \otimes [2]=[4] \oplus [2] \oplus [0]$:
\bes{
&1+x \left[{\blue 1+ \chi^{SU(2)}_{[2]} (f)} \right]  +x^2\Big[ (g''+1) (w'+w'^{-1}) \chi^{SU(2)}_{[4]} (f)   \\
& \quad  + g'' (w'+w'^{-1}) +(g''+2) (\chi^{SU(2)}_{[4]} (f)+ \chi^{SU(2)}_{[2]} (f)+1) +1  \\
& \quad - {\red \chi^{SU(2)}_{[2]} (f)} -  {\blue (\chi^{SU(2)}_{[2]} (f)+1)} \Big]+\ldots
}
where the contribution of the $\CN=3$ flavour currents is denoted in blue and the the contribution of the $\CN=3$ extra supersymmetry current is written in red.  Since there are 3 of the latter, we conclude that supersymmetry gets enhanced from $\CN=3$ to $\CN=3+3=6$.

\subsubsection{$SO(4)_2 \times  USp(4)_{-1}$}
The index of theory II(L): $SO(4)_2 \times  USp(4)_{-1}$ can be written as
\bes{ \label{indIIL}
&1+x\left[1+\zeta \chi^{SU(2)}_{[2]}(f) \right] + x^2\Big[\left( 1+2(1+ \chi) +(1+\chi) \zeta \right) \chi^{SU(2)}_{[4]}(f) \\
& \qquad +(1+\chi)\zeta \chi^{SU(2)}_{[2]}(f) + \left( 1+2(1+ \chi) \right)  - (1-\chi)\zeta - \chi^{SU(2)}_{[2]}(f) \Big] + \ldots~.
}
with the unrefined index given by \eref{indIIRunref}. The indices \eref{indIIR} and \eref{indIIL} can be matched as follows:
\bes{
[ \eref{indIIR} ](u=v=f, \, w'=1, \, g'' =\chi) = [\eref{indIIL}](f, \chi, \zeta = 1)~.
}
In other words, the $SU(2)_f$ flavour symmetry of theory II(L) is identified with the diagonal subgroup of the flavour symmetry $SU(2)_u \times SU(2)_v$ of theory II(R).  The $(Z^{[0]}_2)_{g''}$ zero-form symmetry of theory II(R) is identified with the zero-form charge conjugation symmetry of theory II(L).  However, the $U(1)^{[0]}_{w'}$ zero-form symmetry of theory II(R) is not manifest in theory II(L), whereas the magnetic symmetry $(\BZ^{[0]}_2)_\zeta$ of theory II(L) is not manifest in theory II(R).

\subsection{$O(2N)_2 \times  USp(2N)_{-1}   \,\, \leftrightarrow \,\, U(N)_4 \times U(N)_{-4}$}
In this subsection, we consider the well-known duality between the following two theories:
\bes{
\text{I(L):}~ O(2N)_2 \times  USp(2N)_{-1}   \quad &\leftrightarrow \quad  \text{I(R):}~ U(N)_4 \times U(N)_{-4}~.
}

Theory I(L) can be obtained from theory II(L) by gauging the zero-form charge conjugation symmetry of the latter.  At the level of the index, this can be done by summing over $\chi \in \{ -1, +1 \}$.  As a result, we are left with the fugacity $f$ for the $SU(2)_f$ flavour symmetry and the fugacity $\zeta$ for the zero-form $\BZ_2$ magnetic symmetry. 

On the other hand, theory I(R) can be obtained from theory II(R) by gauging the zero-form symmetry $(\BZ^{[0]}_{\GCD(N,2)})_{g''}$ of the latter.  In particular, given the index of theory II(R) written in terms of $u$, $v$, $w'$ and $g''$, where $(g'')^{\GCD(N,2)} =1$, we are summing over $g'' \in \{e^{2 \pi i j/\GCD(N,2)} \,|\, j=0,1, \ldots, \GCD(N,2)-1  \}$.  As a result, we are left with the fugacities $u$ and $v$ for the $SU(2)_u \times SU(2)_v$ flavour symmetry and the fugacity $w'$ for the $U(1)$ topological symmetry.

\subsubsection{The case of $N=1$}
The $\CN=2$ index for theory I(R): $U(1)_4 \times U(1)_{-4}$ is 
\bes{ \label{indexIRN1}
&1+x \left[ \chi^{SU(2)}_{[1]} (u)\chi^{SU(2)}_{[1]} (v) \right] +x^2\Big[ w' \chi^{SU(2)}_{[4]} (v)+w'^{-1}\chi^{SU(2)}_{[4]} (u) \\
& \qquad +\chi^{SU(2)}_{[2]} (u) \chi^{SU(2)}_{[2]} (v)  - \chi^{SU(2)}_{[2]} (u)-  \chi^{SU(2)}_{[2]} (v) -2\Big]+\ldots
}
In order to write this in terms of the $\CN=3$ index, we set $u=v=f$ and use the tensor product decomposition $[2] \otimes [2] = [4]\oplus[2]\oplus[0]$:
\bes{ 
&1+x \left[ {\blue 1+\chi^{SU(2)}_{[2]}(f)}  \right] +x^2\Big[ (w'+w'^{-1})  \chi^{SU(2)}_{[4]} (f) \\
& \qquad +\chi^{SU(2)}_{[4]} (f)+ \chi^{SU(2)}_{[2]} (f)+1  - {\blue ( \chi^{SU(2)}_{[2]} (f)+ 1 )} -  {\red \chi^{SU(2)}_{[2]} (f)  }-1\Big]+\ldots~,
}
where the blue terms denote the contribution of the $\CN=3$ flavour currents in $SU(2)_f \times U(1)_{w'}$, and the red terms denote the contribution of the $\CN=3$ extra supersymmetry current.  Since there are three of the latter, we conclude that supersymmetry gets enhanced from $\CN=3$ to $\CN=6$, as expected.  The last $-1$ term at order $x^2$ is the contribution of the current of the $U(1)_D$ symmetry, discussed below \eref{indexIIRb}.

The operators contributing at order $x$ of \eref{indexIRN1} correspond to 
\bes{ M_{\alpha \alpha'} =A_\alpha B_{\alpha'}~.}  
Those contributing to the positive terms at order $x^2$ (\ie~ $\CN=2$ preserving marginal operators) are gauge invariant dressed monopole operators and the square of $M$:
\bes{
w'^{-1}\chi^{SU(2)}_{[4]} (u):  &\quad (\CM_{-1})_{\alpha_1 \cdots \alpha_4} =T_{\{-1;-1\}}A_{\alpha_1}A_{\alpha_2}A_{\alpha_3}A_{\alpha_4}~, \\ 
w' \chi^{SU(2)}_{[4]} (v): &\quad (\CM_{+1})_{\alpha'_1 \cdots \alpha'_4} =T_{\{+1;+1\}} B_{\alpha'_1}B_{\alpha'_2}B_{\alpha'_3}B_{\alpha'_4} ~, \\ 
\chi^{SU(2)}_{[2]} (u) \chi^{SU(2)}_{[2]} (v):  &\quad \CQ_{\alpha \beta \alpha' \beta'} =(A_\alpha A_\beta) (B_{\alpha'} B_{\beta'}) = M_{\alpha \alpha'} M_{\beta \beta'}~.
}

On the other hand, the index of theory I(L): $O(2)_{2} \times USp(2)_{-1}$ can be obtained by summing over $\chi \in \{-1,1\}$ in \eref{indIILN1zc}
\bes{ \label{indexILN1}
&1+ x \left[ 1 +\zeta' \, \chi^{SU(2)}_{[2]}(f) \right]  +x^2\Big[ (\zeta'+1)  \chi^{SU(2)}_{[4]} (f) \\
& \qquad +\chi^{SU(2)}_{[4]} (f)+ \chi^{SU(2)}_{[2]} (f)+1  - {\blue ( \chi^{SU(2)}_{[2]} (f)+ 1 )} -  {\red \chi^{SU(2)}_{[2]} (f)  }-\zeta'\Big]+\ldots
}
where $(\zeta')^2=1$.\footnote{The magnetic symmetry of $O(2)_2$ is not $U(1)^{[0]}$, but rather $(\mathbb{Z}_2^{[0]})_{\zeta'}$, see for example \cite[Appendix H]{Cordova:2017vab}. It is interesting to point out that the index is sensitive to this. Indeed, if we compute the index of the $O(2)_{2} \times USp(2)_{-1}$ model treating $\zeta'$ as a $U(1)^{[0]}$ fugacity, that is without imposing $(\zeta')^2=1$ as in \eqref{indexILN1}, we would get fractional coefficients such as $\frac{1}{2}(\zeta'+1/\zeta')$, which is clearly incosistent. This signals that the actual symmetry is $(\mathbb{Z}_2^{[0]})_{\zeta'}$ and by accordingly setting $\zeta'=1/\zeta'$ we get the sensible result \eqref{indexILN1}.}  The indices \eref{indexIRN1} and \eref{indexILN1} can be matched as follows:
\bes{
[\eref{indexIRN1}](u=f, v=f, w'=1) = [\eref{indexILN1}](f, \zeta'=1)~.
}
In other words, the $U(1)^{[0]}_{w'}$ topological symmetry of theory I(R) is not manifest in theory I(L), whereas the magnetic symmetry of theory I(L) is not manifest in theory I(R).  As usual, the $SU(2)_f$ flavour symmetry of theory I(L) is identified with the diagonal subgroup of the $SU(2)_u \times SU(2)_v$ flavour symmetry of theory I(R).

The unrefined indices of theories I(L) and I(R) with $N=1$ are, of course, equal and are given by \cite[Table 1]{Cheon:2012be}:
\bes{
1+4x+11x^2 +12x^3 +25x^4 +12x^5 + \ldots~.
}

\subsubsection{The case of $N=2$}
The $\CN=2$ index for theory I(R): $U(2)_4 \times U(2)_{-4}$ is 
\bes{ \label{indexIRN2}
&1+x \left[ \chi^{SU(2)}_{[1]} (u)\chi^{SU(2)}_{[1]} (v) \right] +x^2\Big[ w' \chi^{SU(2)}_{[4]} (v)+w'^{-1}\chi^{SU(2)}_{[4]} (u) \\
& \qquad +2\chi^{SU(2)}_{[2]} (u) \chi^{SU(2)}_{[2]} (v) +1  - \chi^{SU(2)}_{[2]} (u)-  \chi^{SU(2)}_{[2]} (v) -1\Big]+\ldots
}
As before, the $\CN=3$ index can be obtain by setting $u=v=f$:
\bes{\label{indexIRN2wN3susy}
&1+x \left[ {\blue 1+\chi^{SU(2)}_{[2]}(f)}  \right] +x^2\Big[ (w'+w'^{-1})  \chi^{SU(2)}_{[4]} (f) \\
& \qquad +2\chi^{SU(2)}_{[4]} (f)+2 \chi^{SU(2)}_{[2]} (f)+2 +1  - {\blue ( \chi^{SU(2)}_{[2]} (f)+ 1 )} -  {\red \chi^{SU(2)}_{[2]} (f)  }\Big]+\ldots
}
The operators contributing at order $x$ is
\bes{ \label{mesonIIR}
\chi^{SU(2)}_{[1]} (u)\chi^{SU(2)}_{[1]} (v) : \qquad M_{\alpha \alpha'} = (A_\alpha)^a_i (B_{\alpha'})^i_a~.
}
The marginal operators, which contribute to the positive terms at order $x^2$, are\footnote{Notice that the monopoles $T_{\{\pm1,0;\pm1,0\}}$ are in a representation of the $U(2)\times U(2)$ gauge group that have charges $(\pm1/2,\mp1/2)$ under the $U(1)\times U(1)$ part and that are in the symmetric representation of each of the two $SU(2)$ part.}
\bes{ \label{marginalIR}
\scalebox{0.95}{$
\begin{split}
w'^{-1}\chi^{SU(2)}_{[4]} (u) : \quad &(\CM_{-1})_{\alpha_1 \cdots \alpha_4}=(T_{\{-1,0;-1,0\}})_{(a_1\cdots a_4)}^{(i_1 \cdots i_4)} (A_{\alpha_1})^{a_1}_{i_1} \cdots  (A_{\alpha_4})^{a_4}_{i_4} ~, \\ 
w'\chi^{SU(2)}_{[4]} (v) : \quad &(\CM_{+1})_{\alpha'_1 \cdots \alpha'_4}=(T_{\{+1,0;+1,0\}})^{(a_1\cdots a_4)}_{(i_1 \cdots i_4)} (B_{\alpha'_1})^{i_1}_{a_1} \cdots  (B_{\alpha'_4})^{i_4}_{a_4} ~, \\ 
\chi^{SU(2)}_{[2]} (u) \chi^{SU(2)}_{[2]} (v)+1: \quad  & M_{\alpha \alpha'}M_{\beta \beta'}~, \\
\chi^{SU(2)}_{[2]} (u) \chi^{SU(2)}_{[2]} (v): \quad &\CQ_{\alpha \beta \alpha' \beta'} =(A_\alpha)^a_i (B_{\alpha'})^i_b (A_\beta)^b_j (B_{\beta'})^j_a
\end{split}$}
}
where we comment on the above operators as follows: 
\bi
\item The monopole operators $T_{\{+1,0;+1,0\}}$ and $T_{\{-1,0;-1,0\}}$ transform in the representations $[4_{+4};4_{-4}]$ and $[4_{-4};4_{+4}]$ of the gauge group $U(2) \times U(2)$, respectively.
\item The gauge invariant combinations $M M$ in the third line transform in the representation $\Sym^2 [1;1] = [2;2]+[0;0]$ of $SU(2)_u \times SU(2)_v$. 
\item The gauge invariant combinations $\CQ_{\alpha \beta \alpha' \beta'}$ are subject to the $F$-terms coming from the superpotential of the ABJM theory, namely $\epsilon^{\alpha \beta} \epsilon^{\alpha' \beta'} (A_\alpha)^a_i (B_{\alpha'})^i_b (A_\beta)^b_j (B_{\beta'})^j_a$, and so
\bes{ \label{relationQ}
\epsilon^{\alpha \beta} \CQ_{\alpha \beta \alpha' \beta'}  =0 ~, \qquad \epsilon^{\alpha' \beta'} \CQ_{\alpha \beta \alpha' \beta'} =0~.
}
Thus, $\CQ_{\alpha \beta \alpha' \beta'}$ transform under the representation $[2;2]$ of $SU(2)_u \times SU(2)_v$.
\item Note that one could also consider the following gauge invariant combinations:
\bes{
&(A_{\alpha})^{a_1}_{i_1}(A_{\beta})^{a_2}_{i_2}(B_{\alpha'})_{b_1}^{j_1}(B_{\beta'})_{b_2}^{j_2} \epsilon_{a_1 a_2}\epsilon^{i_1 i_2} \epsilon^{b_1 b_2}\epsilon_{j_1 j_2}\\
&= (A_{\alpha})^{a_1}_{i_1}(A_{\beta})^{a_2}_{i_2}(B_{\alpha'})_{b_1}^{j_1}(B_{\beta'})_{b_2}^{j_2} \, \delta^{[b_1}_{a_1} \delta^{b_2]}_{a_2} \, \delta^{[i_1}_{j_1} \delta^{i_2]}_{j_2} \\
&= (A_{\alpha})^{a_1}_{i_1} (B_{\alpha'})_{a_1}^{i_1} (A_{\beta})^{a_2}_{i_2} (B_{\beta'})^{i_2}_{a_2} - (\alpha' \leftrightarrow \beta') \\
&= M_{\alpha \alpha'} M_{\beta \beta'} - (\alpha' \leftrightarrow \beta')
}
and so they are not independent from those in \eref{marginalIR}.
\ei

The index of theory I(L): $O(4)_{2} \times USp(4)_{-1}$ is
\bes{ \label{indexILN2}
&1+ x \left[ 1 +\zeta' \, \chi^{SU(2)}_{[2]}(f) \right]  +x^2\Big[ (\zeta'+1)  \chi^{SU(2)}_{[4]} (f)+2\chi^{SU(2)}_{[4]} (f) \\
& \qquad + (1+\zeta') \chi^{SU(2)}_{[2]} (f)+2+1- {\blue ( \chi^{SU(2)}_{[2]} (f)+ 1 )} -  {\red \chi^{SU(2)}_{[2]} (f)}+1-\zeta'\Big]+\ldots
}
Let us discuss the operators, contributing to order $x$, in this theory.  In the following, $i,j = 1,...,4$ are the $O(4)$ gauge indices; $a,b =1,..., 4$ are the $USp(4)$ gauge indices; and $\alpha, \beta=1,2$ are the $SU(2)_f$ flavour indices. They are
\bes{
1: &\qquad \fm_{[\alpha \beta]} = (A_\alpha)^{i_1}_{a_1} (A_\beta)^{i_2}_{b_2} \delta_{i_1 i_2} J^{a_1 a_2}~, \\
\zeta' \, \chi^{SU(2)}_{[2]}(f) : &\qquad \mathfrak{M}_{(\alpha \beta)} = (T_{\{1,0; 1,0\}})^{(a_1 a_2)}_{(i_1 i_2)} (A_{\alpha})^{i_1}_{a_1} (A_{\beta})^{i_2}_{a_2}
}
where $\fm$ transforms as a singlet under $SU(2)_f$, due to the total antisymmetrisation of the indices $\alpha$ and $\beta$, and $\mathfrak{M}$ transforms as a triplet under $SU(2)_f$, due to the total symmetrisation of the gauge indices in the elementary monopole operator $T_{\{1,0; 1,0\}}$.  
These operators are mapped to the mesons \eref{mesonIIR} of the unitary theory I(R).  Hence, from the perspective of the $\CN=3$ theory, these are the moment map operators of the $U(1) \times SU(2)_f$ symmetry, whose contribution of the currents is denoted in blue.  The contributions in red are instead identified as the $\CN=3$ extra supersymmetry-currents that make $\CN=3$ supersymmetry become $\CN=3+3=6$ supersymmetry\footnote{It is also interesting to analyse this theory from the perspective of $\CN=5$ theory.  For an SCFT with $\CN=5$ (and not higher) supersymmetry, it is necessary that the coefficient of $x$ in the index must be 1 \cite{Evtikhiev:2017heo}, whose contribution comes from the $\CN=5$ stress-tensor multiplet decomposed into one $\CN=2$ multiplet $L \bar{B}_1 [0]_1^{(1)}$ (in the notation of \cite{Cordova:2016emh}). However, for an SCFT with $\CN=6$ (and not higher) supersymmetry, the coefficient of $x$ must be 4 \cite{Evtikhiev:2017heo}, which is the case for \eref{indexILN2}. Observe that the singlet operator $\fm$ is present in any $O(2N)_{2k} \times USp(2N)_{-k}$ theory, where for $k \geq 2$ the theory has $\CN=5$ supersymmetry \cite{Aharony:2008gk}.  We thus conclude that the operator $\fm$ resides in the $\CN=5$ stress-tensor multiplet $B_1[0]_1^{(1,0)}$, whereas the triplet operators $\mathfrak{M}$ reside in the $\CN=5$ extra supersymmetry-current multiplet $B_1[0]_1^{(0,2)}$. The singlet in the tensor product decomposition of $[0,2] \otimes [0,2]$, where each $[0,2]$ is the representation of the latter multiplet under $\mathfrak{so}(5)_R$ symmetry, corresponds to the $U(1)$ global symmetry, which must be present in any $\CN=6$ SCFT \cite{Bashkirov:2011fr}.  This is mapped to the $U(1)_{w'}$ symmetry of the unitary theory I(R).  We thank Oren Bergman for explaning and pointing this out to us.\label{foot:N5}} \footnote{To elucidate further Footnote \ref{foot:N5}, we provide, as a reference, the index of the $O(4)_{4} \times USp(4)_{-2}$ theory, which has $\CN=5$ supersymmetry:
\bes{
1 + 1x + x^2  \left[ 2+ (1+\zeta) \chi^{SU(2)}_{[4]} (f)  - \chi^{SU(2)}_{[2]} (f) \right] + \ldots
}
From $\CN=2$ perspective, the negative term at order $x^2$ indicates the contribution of the $\CN=2$ $SU(2)_f$ flavour currents in the multiplet $A_2 \bar{A}_2 [0]_1^{(0)}$.  From the $\CN=3$ perspective, the three components of $SU(2)_f$ currents split into two parts.  Suppose that we write the character of the adjoint representation of $SU(2)_f$ as $f^2 + 1 +f^{-2}$.  Two components $(f^2, f^{-2})$ of this $SU(2)_f$ symmetry currents are identified as the $\CN=3$ extra SUSY-currents in the multiplet $A_2[0]_1^{(0)}$; this makes $\CN=3$ supersymmetry become $\CN=3+2=5$ supersymmetry.  The remaining component (corresponding to 1) of this $SU(2)_f$ symmetry currents resides in the $\CN=3$ $U(1)$ flavour current multiplet $B_1[0]_1^{(2)}$. The corresponding moment map operator is the singlet operator $\fm$, contributing $+ 1 x$ to the index. Thus, this $U(1)$ symmetry from the $\CN=3$ perspective is identified as the Cartan subalgebra of $SU(2)_f$.
}.

The indices \eref{indexIRN2} and \eref{indexILN2} can be matched as follows:
\bes{
\eref{indexIRN2}[u=f, v=f, w=1] = \eref{indexILN2}[f, \zeta'=1] 
}
The fugacity $\zeta'$ for the magnetic symmetry of theory I(L) cannot be mapped to any fugacity in theory I(R), and so it is not manifest in theory I(R) and should be considered as emergent in theory I(R). Similarly, the $U(1)^{[0]}_w$ zero-form topological symmetry of theory I(R) should be considered as emergent in theory I(L).  As usual, the $SU(2)_f$ flavour symmetry of theory I(L) is identified as a diagonal subgroup of the $SU(2)_u \times SU(2)_v$ flavour symmetry of theory I(R).

The unrefined indices for theories I(L) and I(R) for $N=2$ are, of course, equal and are given by \cite[Table 1]{Cheon:2012be}:
\bes{
1+4x+22x^2 +56x^3 +131x^4 +252x^5+\ldots~.
}

\subsection{$SO(4)_2 \times USp(2)_{-1} \,\, \leftrightarrow \,\, [U(3)_4 \times U(1)_{-4}]/\BZ_2$} \label{sec:SO4USp2}
As pointed out in \cite[Section 3.3]{Tachikawa:2019dvq}, the consistency conditions of the quotient $[U(N+x)_ k \times U(N)_k ]/\BZ_p$ are 
\bes{ \label{consistencyquotient}
\text{$p$ divides $k$} \qquad \text{and} \qquad \text{$\frac{k x}{p^2} \in \BZ$}~.
}
In this section, we take $N=1$, $x=2$, $k=4$ and $p=2$.  Indeed, the theory in question can be obtained by gauging the $\BZ^{[1]}_2$ one-form symmetry of the $U(3)_4 \times U(1)_{-4}$ theory. The index for $[U(3)_4 \times U(1)_{-4}]/\BZ_2$ reads (here we define $w=(w_1w_2)^{\frac{1}{4}}$)
\bes{ \label{indU3U1modZ2}
&1+ x \left[ \chi^{SU(2)}_{[1]} (u )  \chi^{SU(2)}_{[1]} (v) \right] + x^2\Big[ w^{-1} \chi^{SU(2)}_{[4]} (u)+ w \chi^{SU(2)}_{[4]} (v) \\ 
&\quad + \chi^{SU(2)}_{[2]} (u) \chi^{SU(2)}_{[2]} (v) +g(w^{1/2}+w^{-1/2}) - \chi^{SU(2)}_{[2]} (u)-\chi^{SU(2)}_{[2]} (v)-1  \Big] +\ldots~.
}
As before, we can obtain the $\CN=3$ index by setting $u=v=f$ and compute the relevant tensor product decompositions:
\bes{
&1+ x \left[ {\blue 1+\chi^{SU(2)}_{[2]} (f )} \right] + x^2\Big[ (w+w^{-1}) \chi^{SU(2)}_{[4]} (f) + \chi^{SU(2)}_{[4]} (f) +\chi^{SU(2)}_{[2]} (f)+1\\ 
&\qquad  +g(w^{1/2}+w^{-1/2}) - {\blue (\chi^{SU(2)}_{[2]} (f)+1)}-{\red \chi^{SU(2)}_{[2]} (f)}  \Big] +\ldots
}
where the blue terms are the contribution of the $\CN=3$ flavour currents and the red term is the contribution of the $\CN=3$ extra supersymmetry current.  Therefore, $\CN=3$ supersymmetry gets enhanced to $\CN=6$.

As usual, the operators contributing at order $x$ of \eref{indU3U1modZ2} are the mesons,
\bes{ \label{mesonU3U1modZ2}
M_{\alpha \alpha'} = (A_{\alpha})^a (B_{\alpha'})_a
}
where $a,b,c=1,2,3$ are the $U(3)$ gauge indices. As usual, the monopole operators $T_{-} \equiv T_{\left \{-1, 0,0; -1 \right \}}$ and $T_{+} \equiv T_{\left \{+1, 0,0; +1 \right \}}$ transform in the representations $[[0,4]_{-4}; +4]$ and $[[4,0]_{+4}; -4]$ of the gauge symmetry $U(3) \times U(1)$ respectively, where $[0,4]_{- 4}$ and $[4,0]_{+4}$ are from the 4th symmetric power of the antifundamental and fundamental representation of $U(3)$ respectively.

We can write down the marginal operators, contributing to the positive terms at order $x^2$, as follows: 
\bes{ \label{marginalU3U1modZ2}
\scalebox{0.9}{$
\begin{split}
w^{-1} \chi^{SU(2)}_{[4]} (u): & \qquad (\CM_{-1})_{\alpha_1 \cdots \alpha_4} = (T_{-})_{(a_1 a_2 a_3 a_4)} (A_{\alpha_1})^{a_1} \cdots (A_{\alpha_4})^{a_4} \\ 
w \chi^{SU(2)}_{[4]} (v): & \qquad (\CM_{+1})_{\alpha'_1 \cdots \alpha'_4} = (T_{+})^{(a_1 a_2 a_3 a_4)} (B_{\alpha_1})_{a_1} \cdots (B_{\alpha_4})_{a_4} \\
\chi^{SU(2)}_{[2]} (u) \chi^{SU(2)}_{[2]} (v): &\qquad \CQ_{\alpha \beta \alpha' \beta'} = M_{\beta \alpha'} M_{\alpha \beta'}
\end{split} $}
}
Moreover, there are marginal operators, associated with the terms $g w^{\pm 1/2}$ at order $x^2$ in the index, that involve monopole operators $T_{\pm \frac{1}{2}} \equiv T_{\pm \left \{+\frac{1}{2}, +\frac{1}{2} ,-\frac{1}{2}; \frac{1}{2} \right \}}$, arising from the $\BZ_2$ discrete quotient.  Here $\CQ_{\alpha \beta \alpha' \beta'}$ is defined as in \eref{marginalIR} with the absence of the indices $i, j$, and $\CM_\pm$ are the gauge invariant dressed monopole operators.

The index of $SO(4)_2 \times USp(2)_{-1}$ reads
\bes{ \label{indSO4USp4}
1&+x \left[ 1+ \zeta \chi^{SU(2)}_{[2]}(f) \right] \\
&+ x^2 \left[ (\zeta+2) \chi^{SU(2)}_{[4]}(f)+  \zeta \chi +\chi +1 -\zeta-  \chi^{SU(2)}_{[2]}(f)\right]+\ldots~.
}
The indices \eref{indU3U1modZ2} and \eref{indSO4USp4} can be matched as follows:
\bes{
[\eref{indU3U1modZ2}](u=f,v=f,w=1,g=\chi)= [\eref{indSO4USp4}](f,\zeta=1,\chi) ~.
}
The flavour symmetry $SU(2)_f$ of the orthosymplectic theory can be identified with the diagonal subgroup of the flavour symmetry $SU(2)_u \times SU(2)_v$ of the unitary theory.  The $(\BZ^{[0]}_2)_g$ zero-form symmetry of the unitary theory is identified with the zero-form charge conjugation symmetry of the orthosymplectic theory.  The $U(1)^{[0]}_w$ zero-form topological symmetry of the unitary theory is not manifest in the orthosymplectic theory, whereas the $(\BZ^{[0]}_2)_\zeta $ zero-form magnetic symmetry of the orthosymplectic theory is not manifest in the unitary theory.

The unrefined indices for both theories are equal to
\bes{
[\eref{indU3U1modZ2}](u=1,v=1,w=1,g=1)&= [\eref{indSO4USp4}](f=1,\zeta=1,\chi=1) \\
&= 1 + 4 x + 14 x^2 + 35 x^4+\ldots~.
}

\subsection{$O(4)_2 \times USp(2)_{-1} \,\, \leftrightarrow \,\, U(3)_4 \times U(1)_{-4}$} \label{sec:O4USp2}
The $O(4)_2 \times USp(2)_{-1}$ theory can be obtained from the $SO(4)_2 \times USp(2)_{-1}$ theory by gauging the charge conjugation symmetry of the latter.  Correspondingly, the $U(3)_4 \times U(1)_{-4}$ can be obtained from the $[U(3)_4 \times U(1)_{-4}]/\BZ_2$ theory by gauging the $(\BZ^{[0]}_2)_g$ zero-form symmetry of the latter.

Summing over $g \in \{\pm1\}$ in \eref{indU3U1modZ2}, we obtain the index for the $U(3)_4 \times U(1)_{-4}$ theory as
\bes{ \label{indU3U1}
&1+ x \left[ \chi^{SU(2)}_{[1]} (u )  \chi^{SU(2)}_{[1]} (v) \right] + x^2\Big[ w^{-1} \chi^{SU(2)}_{[4]} (u)+ w \chi^{SU(2)}_{[4]} (v) \\ 
&\quad + \chi^{SU(2)}_{[2]} (u) \chi^{SU(2)}_{[2]} (v) - \chi^{SU(2)}_{[2]} (u)-\chi^{SU(2)}_{[2]} (v)-1  \Big] +\ldots~.
}
The operators are as listed in \eref{mesonU3U1modZ2} and \eref{marginalU3U1modZ2}, except that there are no monopole operators $T_{\pm \frac{1}{2}}$ due to the absence of the discrete $\BZ_2$ quotient.  By the same argument as in the precedent subsection, the index indicates that the theory has $\CN=6$ supersymmetry, in agreement with \cite{Aharony:2008gk}.

Similarly, summing over $\chi \in \{\pm1\}$ in \eref{indSO4USp4} gives the index of the $O(4)_2 \times USp(2)_{-1}$ theory:
\bes{ \label{indO4USp4}
1+x \left[ 1+ \zeta \chi^{SU(2)}_{[2]}(f) \right] + x^2 \left[ (\zeta+2) \chi^{SU(2)}_{[4]}(f)+1 -\zeta-  \chi^{SU(2)}_{[2]}(f)\right]+\ldots~.
}
The indices \eref{indU3U1} and \eref{indO4USp4} can be matched as follows:
\bes{
[\eref{indU3U1}](u=f,v=f,w=1)= [\eref{indO4USp4}](f,\zeta=1) ~.
}
The correspondence between the global symmetries of the $U(3)_4 \times U(1)_{-4}$ theory and the $O(4)_2 \times USp(2)_{-1}$ are as discussed below \eref{indSO4USp4}. The $SU(2)_f$ flavour symmetry of the orthosymplectic theory is identified with the diagonal subgroup of the $SU(2)_u \times SU(2)_v$ of the unitary theory. The $U(1)^{[0]}_w$ zero-form topological symmetry of the unitary theory is not manifest in the orthosymplectic theory, whereas the $(\BZ^{[0]}_2)_\zeta $ zero-form magnetic symmetry of the orthosymplectic theory is not manifest in the unitary theory. The unrefined indices for both theories are equal
\bes{ \label{unrefU3U1O4USp4}
[\eref{indU3U1}](u=1,v=1, w=1) &= [\eref{indO4USp4}](f=1,\zeta=1) \\
&= 1 + 4 x + 12 x^2 + 8 x^3 + 27 x^4+36x^5 +\ldots~,
}
as computed in \cite[(2.8)]{Cheon:2012be}.

\subsection{Circular quivers} \label{sec:circular}
In this subsection, we examine the following duality for $n \geq 3$:
\bes{ \label{dualitycircular}
& [\underbrace{SO(2)_2 \times  USp(2)_{-1} \times \cdots \times SO(2)_2 \times  USp(2)_{-1}}_{\text{$2n$ gauge groups}}]/\BZ_2 \\
& \longleftrightarrow \quad \text{circular quiver \eref{KN}} \\
& \longleftrightarrow \quad \text{circular quiver}~ \underbrace{U(1)_1 \times  U(1)_{-1}  \times \cdots \times U(1)_1 \times  U(1)_{-1}}_{\text{$2n$ gauge groups}}
}
where the theory on the second line, also known as a Kronheimer--Nakajima quiver \cite{kronheimer1990yang}, is described by
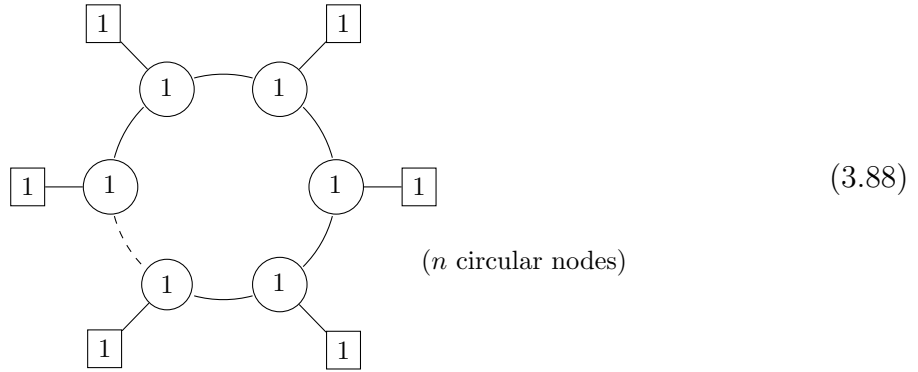
\bea \label{KN}
\begin{tikzpicture}[baseline, align=center,node distance=0.5cm]
\def \n {6}
\def \radius {1.5cm}
\def \margin {15} 
\foreach \s in {1,...,5}
{
  \node[draw, circle] (\s) at ({360/\n * (\s - 2)}:\radius) {{\footnotesize $1_{ }$}};
  \draw[-, >=latex] ({360/\n * (\s - 3)+\margin}:\radius) 
    arc ({360/\n * (\s - 3)+\margin}:{360/\n * (\s-2)-\margin}:\radius);
}
\node[draw, circle] (last) at ({360/3 * (3 - 1)}:\radius) {{\footnotesize $1$}};
\draw[dashed, >=latex] ({360/6 * (5 -2)+\margin}:\radius) 
    arc ({360/6 * (5 -2)+\margin}:{360/6 * (5-1)-\margin}:\radius);
    \node[draw, rectangle,  below right= of 1] (f1) {{\footnotesize $1$}};
\node[draw, rectangle, right= of 2] (f2) {{\footnotesize $1$}};
\node[draw, rectangle, above right= of 3] (f3) {{\footnotesize $1$}};
\node[draw, rectangle, above left= of 4] (f4) {{\footnotesize $1$}};
\node[draw, rectangle,  left= of 5] (f5) {{\footnotesize $1$}};
\node[draw, rectangle,  below left= of last] (f6) {{\footnotesize $1$}};
\draw[-, >=latex] (1) to (f1);
\draw[-, >=latex] (2) to (f2);
\draw[-, >=latex] (3) to (f3);
\draw[-, >=latex] (4) to (f4);
\draw[-, >=latex] (5) to (f5);
\draw[-, >=latex] (last) to (f6);
\node[draw=none] at (4,-1) {{\footnotesize ($n$ circular nodes)}};
\end{tikzpicture}
\eea
This theory is self-mirror, and its Higgs/Coulomb branch describes one $PSU(n) \cong U(n)/U(1)$ instanton on $\BC^2/\BZ_n$ with the holonomy of the gauge field at infinity that brakes $PSU(n)$ into $U(1)^n/U(1)$.

The duality between theories in the second and third lines of \eref{dualitycircular} is well-known and can be seen from the brane system as follows (see \eg~ \cite{Assel:2014awa, Mekareeya:2015bla}).  Theory \eref{KN} can be realised on the worldvolume of a single D3-brane, spanning the directions $0,1,2,3,6$ such that the $6$ direction is compactified with a certain radius, with the presence of $n$ NS5-branes, each spanning the directions $0, 1, 2, 7, 8 ,9$ and one D5-brane, spanning the directions $0, 1, 2, 3, 4 ,5$, at each N5-brane interval  Upon applying the $T^T=-TST$, where $T$ and $S$ are the generators of $SL(2,\BZ)$ such that $S^2=-1$ and $(ST)^3=1$, the NS5 branes remain invariant but each D5-brane turns into a $(1,1)$-brane.  The latter configuration gives rise to the circular quiver in the third line of \eref{dualitycircular}.

\subsubsection*{The case of $n=1$}
In the case of $n=1$, we have seen in Section \ref{sec:IIIN1} that the $[SO(2)_2 \times USp(2)_{-1}]/\BZ_2$ theory flows to a theory of two free hypermultiplets.  This is indeed dual to the special case of \eref{KN} with $n=1$, namely
\bes{
\begin{tikzpicture}[baseline, font= \footnotesize]
\node[draw, circle] (c) at (0,0) {$1$};
\node[draw, rectangle] (r) at (2,0) {$1$};
\draw[-] (c) to (r);
\draw[-] (c) to [out=120,in=-120,looseness=6] (c);
\end{tikzpicture}
}
where the two free hypermultiplets come from the adjoint hypermultiplet of the $U(1)$ gauge group and the elementary monopole operators $T_{\pm1}$. By the above argument and the discussion in Section \ref{sec:IIIN1}, this is also dual to the ABJM theory $U(1)_1 \times U(1)_{-1}$ and $[U(1)_k \times U(1)_{-k}]/\BZ_k$.

\subsubsection*{The case of $n=2$}
The case of $n=2$ requires a separate discussion.  We find that the 
\bes{ \label{SOUSp4nodes}
[SO(2)_2 \times  USp(2)_{-1} \times  SO(2)_2 \times  USp(2)_{-1}]/\BZ_2
}
theory is dual to
\bes{ \label{KN112}
\begin{tikzpicture}[baseline, font= \footnotesize]
\node[draw, circle] (c1) at (0,0) {$1$};
\node[draw, circle] (c2) at (2,0) {$1$};
\node[draw, rectangle] (r) at (4,0) {$2$};
\draw[-] (c1) to[bend left=40] (c2);
\draw[-] (c1) to[bend right=40] (c2);
\draw[-] (c2) to (r);
\end{tikzpicture}
}
which is the Kronheimer-Nakajima quiver whose Higgs/Coulomb branch describes one $PSU(2) \cong U(2)/U(1)$ instanton on $\BC^2/\BZ_2$ with the monodromy that preserves the $PSU(2)$ symmetry.  This is also dual to the following circular quiver:
\bes{ \label{U110m10}
U(1)_1 \times U(1)_0 \times U(1)_{-1} \times U(1)_0
}
The duality between \eref{KN112} and \eref{U110m10} can be realised by applying the action $T^T=-TST$ on the brane system as discussed above.  The $\CN=2$ indices for \eref{KN112} and \eref{U110m10} can be written in terms of the fugacity $d$ of the $U(1)_d$ symmetry and the fugacities $p_1, p_2, p_3, p_4$ of the $SU(2)^4$ global symmetry as follows:
\bes{ \label{indexKN112}
&1+ x \left[d^{-2} \sum_{i=1}^2 \chi^{SU(2)}_{[2]} (p_i) + d^{2} \sum_{i=3}^4 \chi^{SU(2)}_{[2]} (p_i) \right] \\
& \qquad + x^2 \Bigg[ d^{-4} \sum_{i=1}^2 \chi^{SU(2)}_{[4]} (p_i) + d^{4} \sum_{i=3}^4 \chi^{SU(2)}_{[4]} (p_i)\\
& \qquad \qquad +  d^{-4} \chi^{SU(2)}_{[2]} (p_1)  \chi^{SU(2)}_{[2]} (p_2)+  \chi^{SU(2)}_{[2]} (p_1)  \chi^{SU(2)}_{[2]} (p_4)   \\
&\qquad  \qquad+  d^{4} \chi^{SU(2)}_{[2]} (p_3)  \chi^{SU(2)}_{[2]} (p_4) +  \chi^{SU(2)}_{[2]} (p_3)  \chi^{SU(2)}_{[2]} (p_2)\\
&\qquad  \qquad - \left(\sum_{i=1}^4 \chi^{SU(2)}_{[2]} (p_i) \right)-2  \Bigg] +\ldots
}
where the origin of the each $U(1)_d \times SU(2)_{p_i}$ in each theory is as follows. 

For \eref{KN112}, $U(1)_d$ is identified with the axial symmetry that assigns charges $-1$ to each chiral multiplet and $+2$ to the scalar fields in the vector multiplet, $SU(2)_{p_1}$ can be identified with the flavour symmetry that exchanges the two bifundametal hypermultiplets, $SU(2)_{p_2}$ can be identified with the flavour symmetry of the two fundamental hypermultiplets denoted by the square node, $SU(2)_{p_3}$ can be identified with the enhanced $U(1)$ topological symmetry of the left gauge node, and $SU(2)_{p_4}$ can be identified with the enhanced $U(1)$ diagonal subgroup of the $U(1)\times U(1)$ symmetry of the left and right gauge nodes.\footnote{Note that the monopole operators $T_{\{1;0\}}$ and $T_{\{1;1\}}$ have $R$-charge $1$.  The former corresponds to $SU(2)_{p_3}$ symmetry and the latter corresponds to the $SU(2)_{p_4}$ symmetry.}.  Since the theory is self-mirror, the index is invariant under the simultaneous exchange of $d \leftrightarrow d^{-1}$ and $(p_1, p_2) \leftrightarrow (p_3, p_4)$

For \eref{U110m10}, let us label each node to be 1 to 4 from left to right, so that nodes 2 and 4 have zero CS levels. Let $w_1, \ldots, w_4$ be fugacities for topological symmetries of node $1$ to $4$ and let $c_i, c_i^{-1}$ to be the fugacities for the $U(1)$ symmetry that gives charge $+1$ and $-1$ to the chiral multiplets $Q_i, \tQ_i$ carrying gauge charges $(1, -1), (-1,1)$ between the $i$-th and the $(i+1)$-th nodes.  Then, we have the following fugacity maps:
\bes{
p_1^2 = w_4, \quad p_2^2= \frac{c_1 c_2}{w_1 w_2 w_3 w_4}~, \quad p_3^2 = c_3 c_4 (w_1 w_2w_3w_4)~, \quad p_4^2 =w_2~.
}
In other words, the $U(1)$ topological symmetries of the two nodes with zero CS levels get enhanced to $SU(2)$.  The operators associated with the currents of the $SU(2)_{p_2}$ and $SU(2)_{p_3}$ flavour symmetries are, respectively, the dressed monopole operators:
\bes{
T_{\{-1;-1;-1;-1\}}Q_1Q_2~, \qquad  T_{\{+1;+1;+1;+1\}}\tQ_1\tQ_2~, \\
T_{\{+1;+1;+1;+1\}}Q_3Q_4~, \qquad  T_{\{-1;-1;-1;-1\}}\tQ_3\tQ_4~.
}
For \eref{U110m10}, the $U(1)_d$ symmetry assigns the charges $-1$, $-1$, $+1$, $+1$ to the $(Q_1, \tQ_1)$, $(Q_2, \tQ_2)$, $(Q_3, \tQ_3)$, $(Q_4, \tQ_4)$, respectively.  Matching of the unrefined indices of theories \eref{KN112} and \eref{U110m10} is demonstrated in \cite[(5.3)]{Gang:2011xp}:
\bes{
1 + 12x + 42x^2 + 48x^3 + 115x^4+\ldots~.
}
The $\CN=3$ indices of \eref{KN112} and \eref{U110m10} can be obtained from \eref{indexKN112} by setting $d=1$.

Let us now discuss the theory \eref{SOUSp4nodes}. As usual, not all symmetries of the unitary quivers \eref{KN112} and \eref{U110m10} are manifest in the orthosymplectic quiver \eref{SOUSp4nodes}.  The index of the theory \eref{SOUSp4nodes} can be obtained from \eref{indexKN112} by setting $p_3 = p_2$ and $p_4= p_1$, where the origin of $U(1)_d$, $SU(2)_{p_1}$ and $SU(2)_{p_2}$ can be explained as follows. Let $\zeta_1$ and $\zeta_2$ be the fugacities for the $U(1)$ magnetic symmetries for the first and the third $SO(2)$ gauge group respectively.  Let $g$ be a $\BZ_2$ zero-form symmetry arising from the $\BZ_2$ discrete gauging, so that $g^2=1$.  If we denote the half-hypermultiplets in the bifundametal representations of the gauge groups in \eref{SOUSp4nodes}, from left to right, by $A_1$, $A_2$, $A_3$ and $A_4$, the $U(1)_d$ symmetry assigns the charges $+1$, $-1$, $+1$ and $-1$ to them, respectively.  In this notation, the index can be written as follows
\bes{
1+ x \left[ 2 d^{-2}+ 2 d^2 + g \sum_{s_1, s_2, s_3 =\pm1} d^{2s_1} \zeta_1^{\frac{1}{2} s_2} \zeta_2^{\frac{1}{2} s_3}  \right]+\ldots~.
} 
However, the index does not really depend on $g$, since it can be absorbed into a fugacity for the magnetic symmetry. In particular, $\zeta_{1,2}$ and $g$ are related to the fugacities $p_{1,2}$ as follows:
\bes{
\zeta^{1/2}_1 = g p_1 p_2~, \qquad \zeta^{1/2}_2 = p_1^{-1} p_2~.
}
Using this fugacity map, we obtain \eref{indexKN112} with $p_3 = p_2$ and $p_4= p_1$, as required.

\subsubsection*{The case of $n=3$}
The $\CN=2$ index of the unitary theories in the second and third lines of \eref{dualitycircular} can be written as
\bes{ \label{indexcircularn3}
&1+ x(3 d^2 + 3d^{-2})  + x^\frac{3}{2} \Big[ d^{-3} \Big( p_1p_2p_3 + p^{-1}_1p^{-1}_2p^{-1}_3 + \sum_{i=1}^3 (p_i+ p^{-1}_i)  \Big)\\
& \qquad + d^{3} \Big( w_1w_2w_3 + w^{-1}_1w^{-1}_2w^{-1}_3 +\sum_{i=1}^3 (w_i+ w^{-1}_i) \Big) \Big] \\
& \qquad + x^2 \Big[-3+(6 d^4 + 6 d^{-4}) + d^{-4} \sum_{1\leq i < j\leq 3} (p_i p_j + p^{-1}_i p^{-1}_j) \\
& \qquad + d^{4} \sum_{1\leq i < j\leq 3} (w_i w_j + w^{-1}_i w^{-1}_j)  \Big] +\ldots
}
where the theory has a $U(1)^6 \times U(1)_d$ global symmetry, where the fugacities for $U(1)^6$ are denoted by $p_{1,2,3}$ and $w_{1,2,3}$. The $\CN=3$ index can be obtained by setting $d=1$.  

For the theory \eref{KN} with $n=3$, the $U(1)_d$ symmetry corresponds to the axial symmetry that gives that assigns charges $-1$ to each chiral multiplet and $+2$ to the scalar fields in the vector multiplet; the fugacities $w_{1,2,3}$ correspond to the $U(1)^3$ topological symmetry; and the fugacities $p_{1,2,3}$ correspond to the $U(1)^3$ flavour symmetry.  Since the theory is self-mirror, the index is invariant under the simultaneous exchange of $d \leftrightarrow d^{-1}$ and $(p_1, p_2, p_3) \leftrightarrow (w_1,w_2,w_3)$.  To specify our parametrisation of $p_1$, $p_2$, $p_3$, let us first define $c_i, c_i^{-1}$ to be the fugacities for the $U(1)$ symmetry that gives charge $+1$ and $-1$ to the chiral multiplets $Q_i, \tQ_i$ carrying gauge charges $(1, -1), (-1,1)$ between the $i$-th and the $(i+1)$-th gauge nodes, and let $f_i, f_i^{-1}$ be the flavour charges of the fundamental chiral multiplets carrying gauge charge $-1$ and $+1$ under the $i$-th gauge node.  Then, $p_{1,2,3}$ are related to these fugacities as
\bes{
p_1 = f_1 \, c_1 \, f_{2}^{-1},\qquad p_2 = f_2 \, c_2 \, f_{3}^{-1},\qquad p_3 = f_3 \, c_3 \, f_{1}^{-1}~.
}

For the theory on the third line of \eref{dualitycircular}, namely the circular unitary quiver with alternating CS levels, we label the nodes as $1, \ldots, 6$ from left to right. The $U(1)_d$ assigns alternating charges $(-1)^{i+1}$ to the chiral multiplets $(Q_i, \tQ_i)$ in the bifundamental representation of the $i$-th and the $(i+1)$-th gauge nodes.  Let us define $c_i$ (with $i=1,2, \ldots, 6$) as above.  Then, $p_{1,2,3}$ are related to these fugacities as
\bes{
\begin{array}{lll}
p_1= \frac{c_1}{w_1 w_2}~, &\quad p_2= \frac{c_3}{w_3 w_4}~, &\quad p_3= \frac{c_5}{w_5 w_6} \\
w_1= c_2 w_2 w_3~, &\quad w_2 = c_4 w_4 w_5~, &\quad w_3 = c_6 w_6 w_1~.
\end{array}
}
where $w_i$ (with $i=1, 2,\ldots, 6$) the topological symmetry associated with the $i$-th node.

As usual, not all symmetries of these unitary quivers are manifest in the orthosymplectic quiver in the first line of \eref{dualitycircular}.  In fact, the index of the latter can be obtained from \eref{indexcircularn3} by setting $w_i = p_i$, with $i=1,2,3$.  Indeed, if we denote by $\zeta_1, \zeta_2, \zeta_3$ the $U(1)^3$ magnetic symmetry associated with each $SO(2)$ gauge group from left to right, we then have the fugacity map
\bes{
\zeta_1^{1/2} =g p_2^{1/2} p_3^{1/2}~, \qquad \zeta_2^{1/2} = p_1^{1/2} p_2^{1/2}~, \qquad \zeta_3^{1/2} = p_1^{1/2} p_3^{1/2}~,
}
where $g$ is the fugacity associated with a $\BZ_2$ zero-form symmetry associated with the $\BZ_2$ discrete quotient in the first line of \eref{dualitycircular} such that $g^2=1$.  We emphasise that $g$ can be absorbed in a redefinition of a fugacity of the magnetic symmetry and so the index does not really depend on $g$.  The $U(1)_d$ symmetry assigns the charges $(-1)^{i+1}$ to the half-hypermultiplets $A_i$, with $i=1, \ldots, 6$, in the bifundamental representation of $SO(2) \times USp(2)$ from left to right in the circular quiver in the first line of \eref{dualitycircular}.

This discussion can be generalised in a straightforward manner to the cases of $n>3$.

\subsection{$(S)O(2N+1)_{2} \times USp(2N)_{-1}$ and $[U(N+1)_4\times U(N)_{-4}](/\BZ_2)$} \label{sec:SOodd}
In this subsection, we demonstrate that the indices of the following four theories are equal:
\bes{ \label{fouroddtheories}
O(2N+1)_{2} \times USp(2N)_{-1} &\qquad  U(N+1)_4\times U(N)_{-4} \\
SO(2N+1)_{2} \times USp(2N)_{-1} &\qquad  [U(N+1)_4\times U(N)_{-4}]/\BZ_2
}

The duality of the theories in the first line were pointed out in \cite{Aharony:2008gk}.  The one-form symmetry of each theory in the first line is $\BZ_2$, which can be realised as follows.  For the orthosymplectic quiver in the first line, the $O(2N+1)$ and $USp(2N)$ gauge groups both have a $\BZ_2$ centre and the bifundamental matter screens a diagonal combination, so we are left with one $\BZ_2$ centre symmetry. For the unitary quiver in the first line, namely $U(N+1)_4\times U(N)_{-4}$, the presence of the $\BZ_2$ one-form symmetry was pointed out in \cite[Section 3.3]{Tachikawa:2019dvq}.

The theories in the second line arise from gauging the $\BZ_2$ one-form symmetries of the theories on the first line. This is consistent because the conditions \eref{consistencyquotient} are satisfied. We thus expect that the theories in the second line are also dual to each other.  However, what is surprising is that all of the four theories have the same indices.  Let us demonstrate this point as follows.

We first provide an argument to show that the indices of the $U(N+1)_4\times U(N)_{-4}$ theory and the $[U(N+1)_4\times U(N)_{-4}]/\BZ_2$ theory are equal.  We emphasise that, in each of these theories, there is an overall $U(1)$ that does not act on matter fields.  Upon integrating over such a $U(1)$ fugacity in the index, we obtain a delta-function which imposes the following condition that the magnetic fluxes of the $U(N+1)$ gauge group, $m^{(i)}_{L}$, with $i=1, \ldots, N+1$, and those of the $U(N)$ gauge group, $m^{(j)}_R$, with $j=1, \ldots, N$:
\bes{ \label{condfluxfrac}
\sum_{i=1}^{N+1} m^{(i)}_{L} = \sum_{j=1}^{N} m^{(j)}_{R}  
}
In the $[U(N+1)_4\times U(N)_{-4}]/\BZ_2$ theory, we have to sum over the fluxes 
\bes{(m^{(1)}_{L}, \ldots, m^{(N+1)}_{L}; m^{(1)}_{R}, \ldots, m^{(N)}_{R}) \in (\BZ + p/2)^{2N+1} }
and sum over $p \in \{0, 1\}$, whereas in $U(N+1)_4\times U(N)_{-4}$ theory there is a contribution only from the $p=0$ sector.  Observe that, for $p=1$, if one of the two sides of \eqref{condfluxfrac} is half-integral the other is integral.\footnote{This argument can be generalised to any theory of the form $[U(N+x)_{2k} \times U(N)_{-2k}]/\BZ_2$ with $x$ odd.  In such a theory, the $\BZ_2$ zero-form symmetry arising from the discrete gauging acts trivially.}  This means that there is no contribution from the $p=1$ sector to the index, since it is forbidden by \eref{condfluxfrac}.  As a result, the index of the $[U(N+1)_4\times U(N)_{-4}]/\BZ_2$ is the same as that of the $U(N+1)_4\times U(N)_{-4}$ theory.  We see that the $\BZ_2$ zero-form symmetry arises from the $\BZ_2$ discrete gauging of the former theory acts trivially on the theory and hence it is an unfaithful symmetry. This leads us to conclude that the $\BZ_2$ one-form symmetry of the $U(N+1)_4\times U(N)_{-4}$ theory also acts trivially on the line operators.

Similarly, we can provide an argument to show that the zero-form charge conjugation symmetry of the $SO(2N+1)_2 \times USp(2N)_{-1}$ theory acts trivially on the theory and hence it is unfaithful.  The index of this theory can be written as
\begin{align}
&\mathcal{I}_{SO(2N+1)_{2k}\times USp(2N')_{k'}} \nn\\
&=\frac{1}{2^N N!}\sum_{\vec{m}\in\mathbb{Z}^{N}}\oint\prod_{a=1}^N\frac{\udl{z_a}}{2\pi iz_a}\prod_{a=1}^Nx^{-|m_a|}(1-\chi(-1)^{m_a}x^{|m_a|}z_a^{\pm1})\nn\\
&\times\prod_{a<b}^Nx^{-|\pm m_a+m_b|}(1-(-1)^{\pm m_a\pm m_b}x^{|\pm m_a\pm m_b|}z_a^{\pm1}z_b^{\pm1})\prod_{a=1}^Nz_a^{2km_a}\zeta^{m_a}\nn\\
&\times\frac{1}{2^{N'} {N'}!}\sum_{\vec{n}\in\mathbb{Z}^{N'}}\oint\prod_{i=1}^{N'}\frac{\udl{u_i}}{2\pi iu_i}\prod_{i=1}^{N'}x^{-|n_i|}(1-(-1)^{n_i}x^{|n_i|}u_i^{\pm2})\nn\\
&\times\prod_{i<j}^{N'}x^{-|\pm n_i+n_j|}(1-(-1)^{\pm n_i\pm n_j}x^{|\pm n_i\pm n_j|}u_i^{\pm1}u_j^{\pm1})\prod_{i=1}^{N'}u_i^{2k'n_i}\nn\\
&\times\prod_{i=1}^{N'}x^{\frac{-|u_i|}{2}}\frac{\left((-1)^{n_i}\chi\,u^{\mp 1}\,f^{\mp 1}\,x^{\frac{3}{2}+|n_i|};x^2\right)_\infty}{\left((-1)^{n_i}\chi\,u^{\pm 1}\,f^{\pm 1}\,x^{\frac{1}{2}+|n_i|};x^2\right)_\infty}\nn\\
&\times\prod_{a=1}^N\prod_{i=1}^{N'}x^{\frac{-|\pm z_a+u_i|}{2}}\frac{\left((-1)^{\pm m_a+n_i}z_a^{\mp1}\,u^{\mp 1}\,f^{\mp 1}\,x^{\frac{3}{2}+|\pm m_a+n_i|};x^2\right)_\infty}{\left((-1)^{m_i}z^{\pm1}\,u^{\pm 1}\,f^{\pm 1}\,x^{\frac{1}{2}+|\pm m_a+n_i|};x^2\right)_\infty}\,,
\end{align}
where $f$ is the fugacity for the $SU(2)_f$ flavour symmetry, $\zeta$ is the fugacity for the $(\mathbb{Z}_2^{[0]})_\zeta$ topological symmetry satisfying $\zeta^2=1$ and $\chi$ is the fugacity for the zero-form charge conjugation symmetry.  In the problem at hand, we take $N'=N$, $k=1$ and $k'=-1$, but the following argument holds for general $N$, $k$ and $k'$.  We claim that the charge conjugation symmetry can be re-absorbed with a gauge transformation. This can be seen in the index from the fact that if we simultaneously rescale
\be\label{eq:rescalechi}
z_a\to\chi z_a,\quad a=1,\cdots,N\qquad u_i\to\chi u_i,\quad i=1,\cdots,N'
\ee
then the fugacity $\chi$ completely disappears from the matrix integral since $\chi$ is a square root of unity $\chi=\mathrm{e}^{i\pi n}$ with $n=0,1$. What is crucial for this to happen is that the CS level of the $SO(2N+1)_{2k}$ group is even.\footnote{Indeed, if we consider the index of $SO(2N+1)_{2k+1}\times USp(2N')_{k'}$, the CS factor $\prod_{a=1}^Nz_a^{(2k+1)m_a}$ would produce odd powers of $\chi$ after the shift \eqref{eq:rescalechi}, thus leaving a non-trivial $\chi$ dependence.}  This observation leads us to conclude that the $\BZ_2$ one-form symmetry of the $O(2N+1)_2 \times USp(2N)_{-1}$ theory also acts trivially on the spectrum of line operators.  Since we can obtain the theories in the second line of \eref{fouroddtheories} from those in the first line by gauging the $\BZ^{[1]}_2$ one-from symmetry in the latter, from the perspective of the $U(N+1)_4 \times U(N)_{-4}$ theory, such gauging removes from the spectrum Wilson lines in representations that are not multiple of $2$ of $(\mathbf{N+1}, \mathbf{N})$. We thus conjecture that there exist only the Wilson lines in the representation $((\mathbf{N+1})^{2 \fm}, \mathbf{N}^{2 \fm})$, with $\fm \geq 1$, in the spectrum of this theory, and so the action of such a $\BZ^{[1]}_2$ one-form symmetry is trivial.  We leave the verification of this statement to future work.

As a final remark, we see that the four theories in \eref{fouroddtheories} seem to be dual to each other, even though the $\BZ_2$ one-form symmetry seems to be present in the theories in the first line of \eref{fouroddtheories}, but not in the theories in second line.  One might ask if there exists a topological field theory that provides the $\BZ_2$ one-form symmetry in the former.  The answer seems to be no.  This is in contrast with, for example, the duality appetiser \cite{Jafferis:2011ns}, which is a duality between the 3d $\CN=2$ $SU(2)_1$ gauge theory with one adjoint chiral multiplet and a free chiral multiplet together with a topological quantum field theory (TQFT) given by $U(1)_{-2}$.  Indeed, the $SU(2)_1$ gauge theory has a $\BZ_2$ one-form symmetry (as it can be seen from the centre of the gauge group), whereas the theory of a free chiral multiplet does not have any one-form symmetry; in this case the $\BZ_2$ one-form symmetry is provided by the TQFT $U(1)_{-2}$.  The latter can be detected by the index by turning on an appropriate background magnetic flux, which is the one associated with the $U(1)$ flavour symmetry, as we demonstrate  in Appendix \ref{app:appetiser}.  However, upon turning on background magnetic fluxes for the theories on the first line of \eref{fouroddtheories}, we are not able to detect the presence of the TQFT that supports the $\BZ_2$ one-form symmetry.  We thus conclude that such a symmetry acts trivially on the spectrum of the line operators.

Let us report the index of the theories \eref{fouroddtheories} when $N=1$.  It turns out that, up to order $x^2$, those of the unitary theories are given by \eref{indU3U1}, and those of the orthosymplectic theories are given by \eref{indO4USp4}.  Note, however, that from order $x^5$ onwards, they are different; see \cite[Table 1]{Cheon:2012be} for the unrefined indices of these theories:
\bes{
\eref{fouroddtheories}_{N=1}: &\qquad 1+4x+12x^2 +8x^3 +27x^4 +32x^5 +\ldots \\
\eref{unrefU3U1O4USp4}: &\qquad 1+4x+12x^2 +8x^3 +27x^4 +36x^5 +\ldots
}

As a final remark, we also observe that the circular quivers $SO(3)_2 \times USp(2)_{-1} \times SO(3)_2  \times USp(2)_{-1}$ and $O(3)_2 \times USp(2)_{-1} \times O(3)_2  \times USp(2)_{-1}$ have the same indices; up to order $x^2$, they are
\bes{
&1+ x^2 \Big[ (\zeta_1 +\zeta_2 +\zeta_1 \zeta_2 +2) \chi^{SU(2)}_{[4]}(f) +\zeta_1 \zeta_2 +1\\
&\qquad -  (\zeta_1 +\zeta_2 +\zeta_1 \zeta_2 +2)  \chi^{SU(2)}_{[2]}(f)  \Big] +\ldots,
}
where $\zeta_{1,2}$ are fugacities for the magnetic symmetry of the (special)orthogonal gauge groups.  For the theory with special orthogonal gauge groups, the index does not depend on the fugacity for the charge conjugation symmetry.  For reference, we report the unrefined index up to order $x^4$ as follows:
\bes{
1 + 12 x^2 + 4 x^4 +\ldots~.
}

\section{Conclusions and outlook} \label{sec:conclusions}
We have obtained several new dualities between ABJ and related theories, with at least $\CN=6$ supersymmetry, by gauging zero-form or one-form symmetries. We analysed in details the symmetries of these theories and how they are mapped across each duality, paying particular attention on the discrete symmetries. This result is also generalised to a circular quiver with alternating $SO(2)_2$ and $USp(2)_{-1}$ gauge groups and a discrete $\BZ_2$ quotient.

There are several interesting directions for further study. First, it would be interesting to generalise these results to theories with orthosymplectic gauge groups with $\CN=5$ supersymmetry, as well as more general $U(N+x)_k \times U(N)_{-k}$ and $[U(N+x)_k \times U(N)_{-k}]/\BZ_p$ theories with $\CN=6$ supersymmetry. Moreover, regarding the duality involving the circular quivers, it would be nice to find the analog for the higher ranks theories, such as those involving $SO(2N)_2$ and $USp(2N)_{-1}$ gauge groups. Instead, for the theories involving odd orthogonal and symplectic gauge groups, we observed that certain symmetries act trivially on the spectrum of the theories. It would be nice to check and understand this further. It would also be interesting to work out the gravity duals of the theories studied in this paper and understand the dualities from this perspective. 

One last appealing line of possible future investigation would be to analyse along the lines of what was done in this paper possible generalised symmetries of 3d $\mathcal{N}=2$ SCFTs arising from compactification of 5d $\mathcal{N}=1$ SCFTs on Riemann surfaces with fluxes for the global symmetries, a construction that was recently studied in \cite{Sacchi:2021afk,Sacchi:2021wvg}. Various results have been found for generalised symmetries of 5d SCFTs using both field theory and geometric engineering methods, see for example \cite{Morrison:2020ool,Albertini:2020mdx,Closset:2020scj,Bhardwaj:2020phs,BenettiGenolini:2020doj,Apruzzi:2021vcu,Closset:2021lwy}, and it would be interesting to investigate their imprint on the 3d theories resulting from the compactification.\footnote{See also \cite{Eckhard:2019jgg} for a discussion on higher form symmetries of 3d $\mathcal{N}=1,2$ theories arising from the compactification of 6d $(2,0)$ SCFTs on 3-manifolds.}

\acknowledgments
We would like to thank Oren Bergman, Lea E.~Bottini, Alberto Zaffaroni and Gabi Zafrir for useful conversations. The research of MS is partially supported by the University of Milano-Bicocca grant 2016-ATESP0586, by the MIUR-PRIN contract 2017CC72MK003, by the INFN, by the ERC Consolidator Grant \#864828 “Algebraic Foundations of Supersymmetric Quantum Field Theory (SCFTAlg)” and by the Simons Collaboration for the Nonperturbative Bootstrap under grant \#494786 from the Simons Foundation. 

\appendix

\section{3d supersymmetric index conventions} \label{app:index}

In this appendix we summarise our conventions for the 3d supersymmetric index \cite{Bhattacharya:2008zy,Bhattacharya:2008bja, Kim:2009wb,Imamura:2011su, Kapustin:2011jm, Dimofte:2011py} (which coincides with the superconformal index when computed with the superconformal R-charge). This index can be expressed with the following integral form:
\be\label{indexpartitionfunction}
\mathcal{I}(\{\vec{\mu},\vec{n}\})\,=\,\sum_{\vec{m}}\frac{1}{|\mathcal{W}_{\vec{m}}|}\oint_{\mathbb{T}^{\text{rk}G}}\prod_{i=1}^{\text{rk}G}\frac{\udl{z_a}˘}{2\pi i z_a} \mathcal{Z}_{\text{cl}}(\{\vec{z},\vec{m}\})\,\mathcal{Z}_{\text{vec}}(\{\vec{z},\vec{m}\})\,\mathcal{Z}_{\text{mat}}(\{\vec{z},\vec{m}\};\{\vec{\mu},\vec{n}\})\,,
\ee
where we denoted by $\vec{z}$ the gauge fugacities and by $\vec{m}$ the corresponding magnetic fluxes. The integration contour is taken to be the unit circle $\mathbb{T}$ for each integration variable and the prefactor $|\mathcal{W}_{\vec{m}}|$ is the dimension of the Weyl group of the residual gauge symmetry in the monopole background labelled by the configuration of magnetic fluxes $\vec{m}$.  We also use $\{\vec{\mu},\vec{n}\}$ to denote possible fugacities and fluxes for global symmetries, respectively. In the following we discuss the different contributions to the integrand of \eqref{indexpartitionfunction} in some details.

The factor $\mathcal{Z}_{\text{cl}}$ denotes the classical contributions, which can consist of Chern--Simons interactions and, when the gauge group contains some abelian factor, FI interactions. For example, for a $U(N)$ gauge group it takes the form
\bea
\mathcal{Z}^{U(N)}_{\text{cl}}(\{\vec{z},\vec{m}\})=\prod_{a=1}^Nz_a^{km_a}w^{m_a}\,,
\eea
where $k$ is the CS level and $w$ is the fugacity associated with the $U(1)_w^{[0]}$ zero-form topological symmetry. In the main text we consider also $USp(2N)$, $SO(2N)$ and $SO(2N+1)$ gauge groups, for which we use a different normalisation of the CS level and the topological symmetry is either discrete or absent:
\bea
\mathcal{Z}_{\text{cl}}^{USp(2N)}(\{\vec{z},\vec{m}\})&=\prod_{a=1}^Nz_a^{2km_a}\nn\\
\mathcal{Z}^{SO(2N+\epsilon)}_{\text{cl}}(\{\vec{z},\vec{m}\})&=\prod_{a=1}^Nz_a^{2km_a}\zeta^{m_a}\,,
\eea
where for compactness we denoted $SO(2N+\epsilon)$ for $\epsilon=0,1$.
Moreover, $\zeta$ is the fugacity for the zero-form topological symmetry, which is $U(1)_{\zeta}^{[0]}$ for $SO(2)\cong U(1)$, while it is a $(\mathbb{Z}_2^{[0]})_\zeta$ for $SO(2N)$ and $SO(2N+1)$ with $N>1$ so in these cases we have the condition $\zeta^2=1$.

The factor $\mathcal{Z}_{\text{vec}}$ denotes the contribution of 3d $\mathcal{N}=2$ vector multiplets, which takes the following generic form:
\be\label{indvect}
\mathcal{Z}_{\text{vec}}(\{\vec{z},\vec{m}\})\,=\,\prod_{\alpha\in\text{Lie}(G)}x^{-\frac{|\alpha(\vec{m})|}{2}}(1-(-1)^{\alpha(\vec{m})}\vec{z}^{\alpha}x^{|\alpha(\vec{m})|})\,,
\ee
where $\alpha$ are roots in the gauge algebra $\text{Lie}(G)$ and we are using the short-hand notations
\be
\vec{z}^\ga=\prod_{a=1}^{\text{rk}G}z_a^{\ga_a},\quad \ga(\vec{m})=\sum_{a=1}^{\text{rk}G}\ga_am_a,\quad|\alpha(\vec{m})|=\sum_{a=1}^{\text{rk}G}\ga_am_a\,.
\ee
Explicitly for the groups of interest in this paper we have
\bea\label{induspvect}
\mathcal{Z}_{\text{vec}}^{U(N)}(\{\vec{z},\vec{m}\})&=\prod_{a<b}^Nx^{-|m_a-m_b|}(1-(-1)^{m_a-m_b}z_az_b^{-1}x^{|m_a-m_b|})\nn\\
&\times(1-(-1)^{m_a-m_b}z_a^{-1}z_bx^{|m_a-m_b|})\nn\\
\mathcal{Z}_{\text{vec}}^{USp(2N)}(\{\vec{z},\vec{m}\})&=\prod_{a=1}^Nx^{-2|m_a|}(1-(-1)^{2m_a}z_a^2x^{2|m_a|})(1-(-1)^{2m_a}z_a^{-2}x^{2|m_a|})\nn\\
&\times\prod_{a<b}^Nx^{-|m_a+m_b|-|m_a-m_b|}\times(1-(-1)^{m_a+m_b}z_az_bx^{|m_a+m_b|})\nn\\
&\times(1-(-1)^{m_a-m_b}z_a^{-1}z_b^{-1}x^{|m_a-m_b|})(1-(-1)^{m_a-m_b}z_az_b^{-1}x^{|m_a-m_b|})\nn\\
&\times(1-(-1)^{m_a-m_b}z_a^{-1}z_bx^{|m_a-m_b|})\nn\\
\mathcal{Z}_{\text{vec}}^{SO(2N+\epsilon)}(\{\vec{z},\vec{m}\},\chi=+1)&=\left(\prod_{a=1}^Nx^{-|m_a|}(1-(-1)^{m_a}z_ax^{|m_a|})(1-(-1)^{m_a}z_a^{-1}x^{|m_a|})\right)^\epsilon\nn\\
&\times\prod_{a<b}^Nx^{-|m_a+m_b|-|m_a-m_b|}\times(1-(-1)^{m_a+m_b}z_az_bx^{|m_a+m_b|})\nn\\
&\times(1-(-1)^{m_a-m_b}z_a^{-1}z_b^{-1}x^{|m_a-m_b|})(1-(-1)^{m_a-m_b}z_az_b^{-1}x^{|m_a-m_b|})\nn\\
&\times(1-(-1)^{m_a-m_b}z_a^{-1}z_bx^{|m_a-m_b|})\,,
\eea
where again $\epsilon=0,1$. Remember that for $SO(2N+\epsilon)$ we also have a discrete zero-form charge conjugation symmetry $(\mathbb{Z}_2^{[0]})_\mathcal{C}$ whose corresponding fugacity in the index we denote by $\chi$. The above expressions hold for $\chi=+1$, while for $\chi=-1$ we have \cite{Hwang:2011qt,Hwang:2011ht,Aharony:2013kma}
\bea
\mathcal{Z}_{\text{vec}}^{SO(2N)}(\{\vec{z},\vec{m}\};\chi=-1)&=\prod_{a=1}^{N-1}x^{-2|m_a|}(1-(-1)^{2m_a}z_a^2x^{2|m_a|})(1-(-1)^{2m_a}z_a^{-2}x^{2|m_a|})\nn\\
&\times\prod_{a<b}^{N-1}x^{-|m_a+m_b|-|m_a-m_b|}\times(1-(-1)^{m_a+m_b}z_az_bx^{|m_a+m_b|})\nn\\
&\times(1-(-1)^{m_a-m_b}z_a^{-1}z_b^{-1}x^{|m_a-m_b|})(1-(-1)^{m_a-m_b}z_az_b^{-1}x^{|m_a-m_b|})\nn\\
&\times(1-(-1)^{m_a-m_b}z_a^{-1}z_bx^{|m_a-m_b|})\,,\\
\mathcal{Z}_{\text{vec}}^{SO(2N+1)}(\{\vec{z},\vec{m}\};\chi=-1)&=\left(\prod_{a=1}^Nx^{-|m_a|}(1+(-1)^{m_a}z_ax^{|m_a|})(1+(-1)^{m_a}z_a^{-1}x^{|m_a|})\right)^\epsilon\nn\\
&\times\prod_{a<b}^Nx^{-|m_a+m_b|-|m_a-m_b|}\times(1-(-1)^{m_a+m_b}z_az_bx^{|m_a+m_b|})\nn\\
&\times(1-(-1)^{m_a-m_b}z_a^{-1}z_b^{-1}x^{|m_a-m_b|})(1-(-1)^{m_a-m_b}z_az_b^{-1}x^{|m_a-m_b|})\nn\\
&\times(1-(-1)^{m_a-m_b}z_a^{-1}z_bx^{|m_a-m_b|})\,.
\eea
In particular, the expression for $\chi=-1$ in the $SO(2N)$ case is obtained by setting $z_N=1$, $z_N^{-1}=-1$ and $m_N=0$ in the one for $\chi=+1$,\footnote{Notice that such replacement is not possible when $m_a$ are all half-integer fluxes, which is what happens when we take the gauge group (so not just the Lie algebra) to be $SO(2N)/\mathbb{Z}_2$. In other words, for $SO(2N)/\mathbb{Z}_2$ there is no $\chi=-1$ sector and the $(\mathbb{Z}^{[0]}_2)_{\mathcal{C}}$ charge conjugation symmetry acts trivially.} while the expression for generic $\chi$ in the $SO(2N+1)$ can also be written compactly as
\bea
\mathcal{Z}_{\text{vec}}^{SO(2N+1)}(\{\vec{z},\vec{m}\};\chi)&=\prod_{a=1}^Nx^{-|m_a|}(1-(-1)^{m_a}\chi z_ax^{|m_a|})(1-(-1)^{m_a}\chi z_a^{-1}x^{|m_a|})\nn\\
&\times\prod_{a<b}^Nx^{-|m_a+m_b|-|m_a-m_b|}\times(1-(-1)^{m_a+m_b}z_az_bx^{|m_a+m_b|})\nn\\
&\times(1-(-1)^{m_a-m_b}z_a^{-1}z_b^{-1}x^{|m_a-m_b|})(1-(-1)^{m_a-m_b}z_az_b^{-1}x^{|m_a-m_b|})\nn\\
&\times(1-(-1)^{m_a-m_b}z_a^{-1}z_bx^{|m_a-m_b|})\,.
\eea

Finally, we have the contribution of matter fields which come into 3d $\mathcal{N}=2$ chiral multiplets. The contribution of a chiral with $R$-charge $r$ and transforming under a $U(1)$ symmetry with fugacity and flux $z$ and $m$ respectively is
\bes{\label{indchir}
\CZ_{\text{chir}} (z; m ; R) = (x^{1-R} z^{-1})^{|m|/2} \prod_{j=0}^\infty \frac{1-(-1)^m z^{-1} x^{|m|+2-R +2j}}{1-(-1)^m z x^{|m|-R +2j}}~.
}
The full contribution to the index of a set of chirals forming some representation $\mathbf{R}$ and $\mathbf{R}_F$ of the gauge and the flavour symmetry respectively and with $R$-charge $r$ is
\bea
\mathcal{Z}_{\text{mat}}(\{\vec{z},\vec{m}\};\{\vec{\mu},\vec{n}\})&=\prod_{\rho\in\mathbf{R}}\prod_{\rho_F\in\mathbf{R}_F}\CZ_{\text{chir}} (\vec{z}^{\rho}\vec{\mu}^{\rho_F}; \rho(\vec{m})+\rho_F(\vec{n}) ; r)\,,
\eea
where $\rho$ and $\rho_F$ are the weights of $\mathbf{R}$ and $\mathbf{R}_F$ respectively. Notice that a chiral in the adjoint representation of the gauge group and with $R$-charge 1 gives a trivial contribution to the index, so that the index factor for a 3d $\mathcal{N}=4$ vector multiplet actually coincides with the one of an $\mathcal{N}=2$ vector multiplet \eqref{indvect}. Other examples of matter fields that we encountered in the main text are 3d $\mathcal{N}=4$ hypers in the bifundamental of $U(N)\times U(M)$
\bea
\mathcal{Z}^{U(N)\times U(M)}_{\text{mat}}(\{\vec{z},\vec{m}\};\{\vec{\mu},\vec{n}\})&=\prod_{a=1}^N\prod_{b=1}^M \CZ_{\text{chir}} (z_a/w_b; m^{(1)}_a-m^{(2)}_b ; 1/2)\nn\\
&\qquad\quad\times\CZ_{\text{chir}} (w_a/z_b; m^{(2)}_a-m^{(1)}_b ; 1/2)
\eea
and hypers in the bifundamental of $SO(2N+\epsilon)\times USp(2M)$ for $\epsilon=0,1$
\bea
&\mathcal{Z}^{SO(2N+\epsilon)\times USp(2N)}_{\text{mat}}(\{\vec{z},\vec{m}\};\{\vec{\mu},\vec{n}\};\chi=+1)=\nn\\
&\qquad\qquad\qquad=\left(\prod_{b=1}^M\CZ_{\text{chir}} (1/w_b; -m^{(2)}_b ; 1/2)\CZ_{\text{chir}} (w_b; m^{(2)}_b ; 1/2)\right)^\epsilon \nn\\
&\qquad\qquad\qquad\times\prod_{a=1}^N\prod_{b=1}^M\CZ_{\text{chir}} (z_aw_b; m^{(1)}_a+m^{(2)}_b ; 1/2)\CZ_{\text{chir}} (z_a/w_b; m^{(1)}_a-m^{(2)}_b ; 1/2)\nn\\
&\qquad\qquad\qquad\times\CZ_{\text{chir}} (w_a/z_b; m^{(2)}_a-m^{(1)}_b ; 1/2)\CZ_{\text{chir}} (1/z_aw_b; -m^{(1)}_a-m^{(2)}_b ; 1/2)\,,
\eea
where $\vec{z}$are the fugacities of the left node and $\vec{w}$ those of the right node.
The last expression holds again only for $\chi=+1$. The correct contribution for $\chi=-1$ in the case $\epsilon=0$ is obtained by setting $z_N=1$, $z_N^{-1}=-1$ and $m_N=0$, while when $\epsilon=1$ we have the compact expression for generic $\chi$
\bea
&\mathcal{Z}^{SO(2N+1)\times USp(2N)}_{\text{mat}}(\{\vec{z},\vec{m}\};\{\vec{\mu},\vec{n}\};\chi)=\nn\\
&\qquad\qquad\qquad=\prod_{b=1}^M\left(\CZ_{\text{chir}} (\chi/w_b; -m^{(2)}_b ; 1/2)\CZ_{\text{chir}} (\chi w_b; m^{(2)}_b ; 1/2)\right)^\epsilon \nn\\
&\qquad\qquad\qquad\times\prod_{a=1}^N\prod_{b=1}^M\CZ_{\text{chir}} (z_aw_b; m^{(1)}_a+m^{(2)}_b ; 1/2)\CZ_{\text{chir}} (z_a/w_b; m^{(1)}_a-m^{(2)}_b ; 1/2)\nn\\
&\qquad\qquad\qquad\times\CZ_{\text{chir}} (w_a/z_b; m^{(2)}_a-m^{(1)}_b ; 1/2)\CZ_{\text{chir}} (1/z_aw_b; -m^{(1)}_a-m^{(2)}_b ; 1/2)\,.
\eea

\section{Topological sectors and indices: the example of the duality appetiser} \label{app:appetiser}

In Section \ref{sec:SOodd} we mentioned that the 3d index can be useful for detecting the presence of topological sectors in a theory. In this appendix we give some details of this for the case of the duality appetiser of \cite{Jafferis:2011ns}. This is a duality which relates an $SU(2)_1$ gauge theory with one adjoint chiral to the product theory of a free chiral multiplet plus a topological sector consisting of a $U(1)_{-2}$ TQFT. The topological sector was detected in \cite{Jafferis:2011ns} using the $\mathbb{S}^3$ partition function, where it was observed that the $U(1)_f$ symmetry acting on the adjoint chiral on the $SU(2)_1$ side of the duality is mapped to a combination of the R-symmetry and the topological symmetry of the $U(1)_{-2}$ TQFT on the dual side. The topological sector can actually be detected also in the index by turning on a background magnetic flux $m_f$ for the $U(1)_f$ symmetry. The duality is indeed represented by the following identity of indices:
\begin{align}\label{eq:indapp}
&\frac{1}{2}\sum_{m=-\infty}^{+\infty}\oint\frac{\udl{z}}{2\pi iz}z^{2m}x^{-2m}(1-x^{2m}z^{\pm2})(f^2x^{R_\Gp-1})^{-|m_f|}\frac{(f^{-2}\,x^{2-R_\Gp+|2m_f|};x^2)_\infty}{(f^2\,x^{R_\Gp+|2m_f|};x^2)_\infty}\nn\\
&\times(z^{\pm2}f^2x^{R_\Gp-1})^{-|\pm m+m_f|}\frac{(z^{\mp2}\,f^{-2}\,x^{2-R_\Gp+|\mp2 m+2m_f|};x^2)_\infty}{(z^{\pm2}\,f^2\,x^{R_\Gp+|\pm2 m+2m_f|};x^2)_\infty}=\nn\\
&=(f^4x^{2R_\Gp-1})^{-|2m_f|}\frac{(f^{-4}\,x^{2-2R_\Gp+|4m_f|};x^2)_\infty}{(f^4\,x^{2R_\Gp+|4m_f|};x^2)_\infty}\sum_{m=-\infty}^{+\infty}\oint\frac{\udl{z}}{2\pi iz}z^{2m+2m_f}\left(x^{R_\Gp+1}f^2\right)^m\,,
\end{align}
where $f$, $m_f$ are the fugacity and the flux for the $U(1)_f$ global symmetry and $R_\Gp$ is the R-charge of the adjoint chiral field. The identity holds provided that the background flux is quantised as $m_f\in\mathbb{Z}$ and for generic values of $f$ and $R_\Gp$. Notice that the map between the $U(1)_f$ symmetry acting on $\Gp$ and the topological symmetry of the TQFT is compatible with the one found in \cite{Jafferis:2011ns} at the level of the $\mathbb{S}^3$ partition function. The prefactor to the integral on the right hand side is the index of the free chiral, while the remaining integral is the index of the $U(1)_{-2}$ TQFT, which evaluates to
\begin{align}\label{eq:indtopapp}
\sum_{m=-\infty}^{+\infty}\oint\frac{\udl{z}}{2\pi iz}z^{2m+2m_f}\left(x^{R_\Gp+1}f^2\right)^m=\left(x^{R_\Gp+1}f^2\right)^{-m_f}
\end{align}
and, as we anticipated, is trivial if we turn off the background magnetic flux $m_f=0$, while it is non-trivial if we take $m_f\in\mathbb{Z}_{\neq 0}$. In other words, the topological sector is detectable by the index provided that we introduce such background flux.

The identity \eqref{eq:indapp} can be tested by perturbatively expanding both sides in $x$. Taking $R_\Gp=\frac{1}{4}$, which is the value corresponding to the superconformal R-symmetry \cite{Jafferis:2011ns}, we find that the indices of both of the dual theories are for $m_f=0$
\begin{align}
\mathcal{I}_{m_f=0}&=1+f^4x^{\frac{1}{2}}+\left(f^{12}-f^{-4}\right)x^{\frac{3}{2}}+f^8x+(f^{16}-1)x^2+f^{20}x^{\frac{5}{2}}+f^{24}x^3+\nn\\
&+\left(f^{28}-f^{-4}\right)x^{\frac{7}{2}}+(f^{32}-2)x^4+\left(f^{36}-f^4\right)x^{\frac{9}{2}}+\left(f^{40}+f^{-8}\right)x^5+\nn\\
&+f^{44}x^{\frac{11}{2}}+(f^{48}-2)x^6+\mathcal{O}\left(x^{\frac{13}{2}}\right)\,,
\end{align}
which is the same result that was found in eq. (9) of \cite{Jafferis:2011ns}. Notice that $-1$ at order $x^2$, which represents the fermionic superpartner of the $U(1)_f$ conserved current. For non-trivial $m_f$ we find that the two indices still match and that the topological sector \eqref{eq:indtopapp} is crucial for the matching. For example for $m_f=1$ we get
\begin{align}
\mathcal{I}_{m_f=\pm1}=f^{-10}x^{-\frac{1}{4}}+f^{-6}x^{\frac{17}{4}}-f^{-14}x^{\frac{21}{4}}+\mathcal{O}\left(x^{\frac{25}{4}}\right)\,,
\end{align}
but we also checked \eqref{eq:indapp} for several other values of $m_f$ and $R_\Gp$.

\bibliographystyle{ytphys}
\bibliography{ref}

\end{document}